\documentclass[10pt]{iopart}
\usepackage{graphicx}
\usepackage{subfigure}
\expandafter\let\csname equation*\endcsname\relax
\expandafter\let\csname endequation*\endcsname\relax
\usepackage{amsmath}
\usepackage{amssymb}
\usepackage{iopams}
\usepackage{array}
\newcommand {\Be}{\begin{eqnarray*}}
\newcommand {\Ee} {\end{eqnarray*}}
\newcommand {\bey} {\begin{eqnarray}}
\newcommand {\eey} {\end{eqnarray}}
\newcommand{\bit}{\begin{itemize}}      
\newcommand{\eit}{\end{itemize}}
\newcommand{\bfl}{\begin{flusleft}}
\newcommand{\efl}{\end{flusleft}}
\newcommand{\bfr}{\begin{flushright}}
\newcommand{\bc}{\begin{center}}
\newcommand{\ec}{\end{center}}
\newcommand{\ben}{\begin{enumerate}}    
\newcommand{\een}{\end{enumerate}}

\newcommand{\be}{\begin{equation}}
\newcommand{\ee}{\end{equation}}
\def\<{\langle}
\def\>{\rangle}
\begin{document}
%%%%%%%%%%%%%%%%%%%%%%%%%%%%%%%%%%%%%%%%%%%%%%%%%%%%%%%%%%%%%%%%%%%%%%%%%%%
\title[QM/MM-based quantum master equation description of transport in the PE545 complex]
      {Exciton transport in the PE545 complex: insight from atomistic QM/MM-based 
       quantum master equations and elastic network models} 
%--------------------------------------------------------------------------
\author{Sima Pouyandeh$^{1,2,3}$,
        Stefano Iubini$^{3,4}$,
        Sandro Jurinovich$^5$,
        Yasser Omar$^{1,2}$,
        Benedetta Mennucci$^5$,
        Francesco Piazza$^{3,6}$}
\address{$^1$  Instituto de Telecomunica\c{c}\~oes, Physics of Information and Quantum Technologies Group, Portugal \\
         $^2$  Instituto Superior T\'{e}cnico, Universidade de Lisboa, Portugal \\
         $^3$  Centre de Biophysique Mol\'eculaire, (CBM), CNRS UPR 4301, Rue C. Sadron, 45071, Orl\'eans, France \\
         $^4$  Dipartimento di Fisica, Universit\`a di Firenze and INFN, Sezione di Firenze, Via G. Sansone 1, 50019 Sesto 
               F.no (FI), Italy \\
         $^5$  Dipartimento di Chimica e Chimica Industriale, University of Pisa, via G. Moruzzi 13, 56124 Pisa, Italy \\
         $^6$  Universit\'e d'Orl\'eans, Ch\^ateau de la Source, F-45071 Orl\'eans Cedex, France}
\ead{Francesco.Piazza@cnrs-orleans.fr}

%
%%%%%%%%%%%%%%%%%%%%%%%%%%%%%%%%%%%%%%%%%%%%%%%%%%%%%%%%%%%%%%%%%%%%%%%%%%%%%%%%%%%%%%%%%%%%%%%%%%%%%%%%%%%%%%%%%%%%%%%%%%%%%
%
\begin{abstract} 
In this paper we work out a parameterization of the environment noise within the Haken-Strobl-Reinenker (HSR) model
for the PE545 light-harvesting complex based on atomic-level quantum mechanics/molecular 
mechanics (QM/MM) simulations. We use this approach to investigate the role of different auto- and
cross-correlations in the HSR noise tensor, confirming that site-energy autocorrelations (pure dephasing) terms 
dominate the noise-induced exciton mobility enhancement, followed by site energy-coupling cross-correlations 
for specific triplets of pigments. Interestingly, several cross-correlations of the latter kind, together 
with coupling-coupling cross-correlations, display clear low-frequency signatures in their spectral densities 
in the region 30-70 inverse centimeters. These slow components lie at the limits of validity of the HSR 
approach, requiring that environmental fluctuations be faster than typical exciton transfer time scales. 
We show that a simple coarse-grained elastic-network-model (ENM) analysis of the PE545 protein naturally 
spotlights collective normal modes in this frequency range, that represent specific concerted motions of 
the subnetwork of cysteines that are covalenty linked to the pigments.  
This analysis strongly suggests that protein scaffolds in light-harvesting complexes are able 
to express specific collective, low-frequency normal modes providing a fold-rooted blueprint 
of exciton transport pathways. We speculate that ENM-based mixed quantum classical methods,
such as Ehrenfest dynamics, might be promising tools to disentangle the fundamental designing 
principles of these dynamical processes in natural and artificial light-harvesting structures.
\end{abstract} 

%
%
%------------------------------- SOME PACS -------------------------------------------------------------------------------------
%
%  03.65.Yz	    Decoherence; open systems; quantum statistical methods
%  05.40.Jc     Brownian motion
%  05.45.?a     Nonlinear dynamical systems
%  05.45.Yv 	    Solitons 
%  05.60.Gg	    Quantum transport
%  63.20.Pw     Lattice dynamics
%  71.35.-y     Excitons and related phenomena
%  82.20.Wt     Computational modeling; simulation
%  82.39.Rt     Reactions in complex biological systems (see also 87.18.-h Biological complexity)
%  82.70.Dd     Colloids
%  87.15.-v     Biomolecules: structure and physical properties
%               87.15.Ya	    Fluctuations
%               87.15.H-  	Dynamics of biomolecules
%
%--------------------------------------------------------------------------------------------------------------------------------
\vspace{2pc}

\submitto{\PB}

\maketitle
\tableofcontents

%%%%%%%%%%%%%%%%%%%%%%%%%%%%%%%%%%%%%%%%%%%%%%%%%%%%%%%%%%%%%%%%%%%%%%%%%%%%%%%%%%%%%%%%%%%%%%%%%%%%%%%%%%%%%%%%%%%%%%%%%%%%%%%
\section{Introduction}

\begin{figure}%[t!,clip]
\begin{center}
\includegraphics[width=0.7\textwidth]{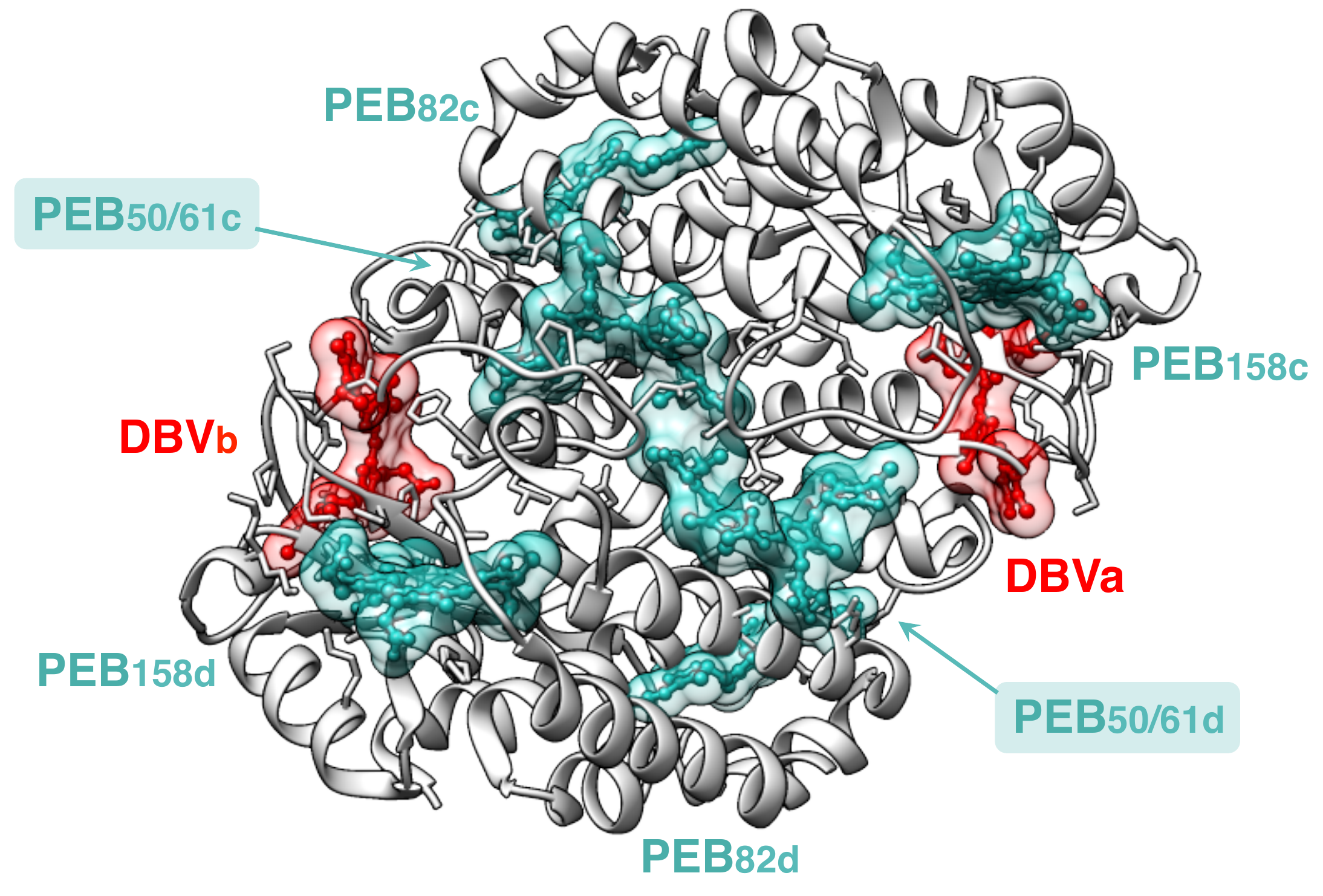}
\includegraphics[width=0.65\textwidth]{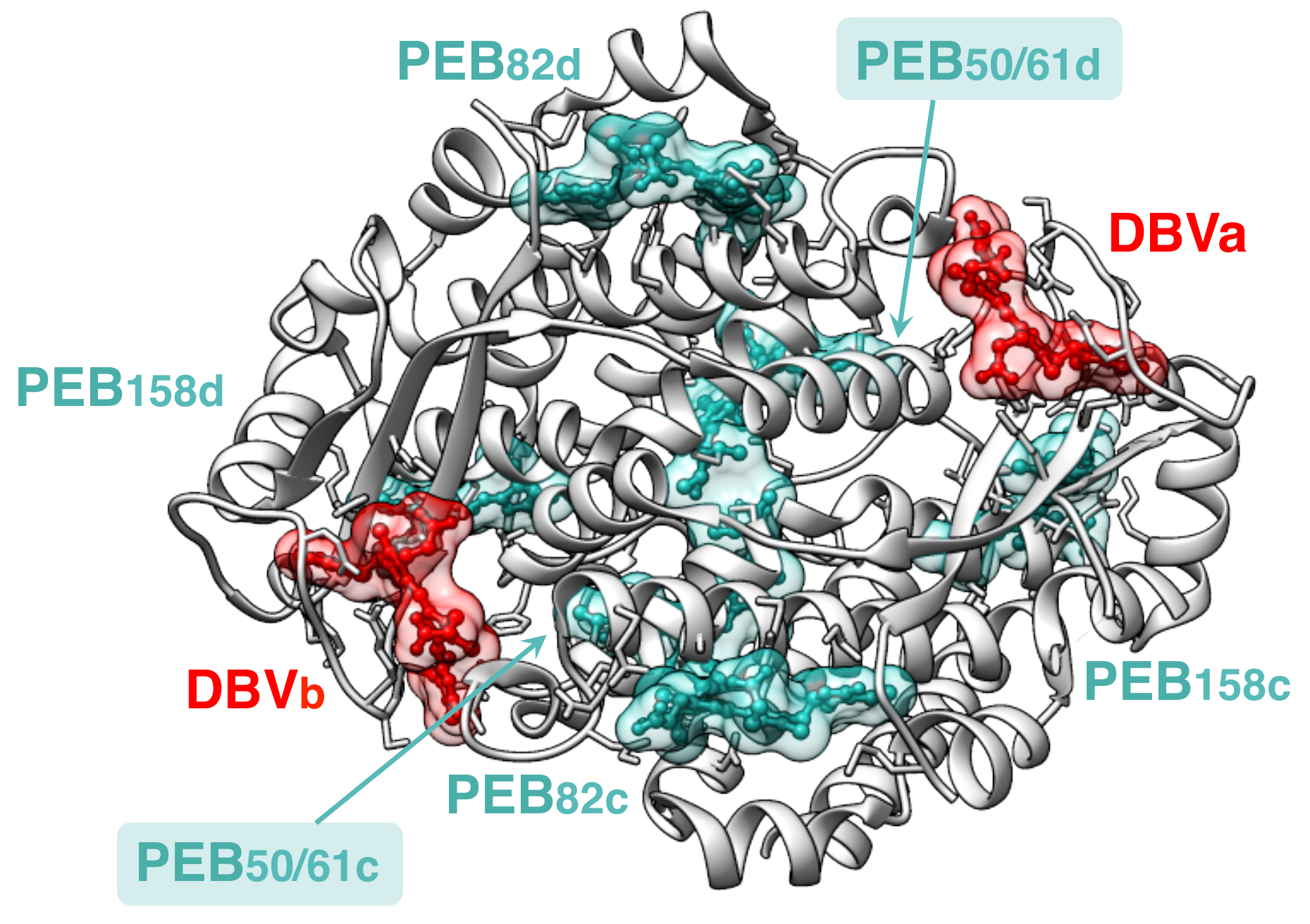}
\end{center}
\caption{The structure of the PE545 light-harvesting complex (dimer) with 
the eight light-absorbing pigments (bilins) shown explicitly. There are two types of bilins in the 
complex:  dihydrobiliverdins (DBV) and phycoerythrobilins (PEB). Both are  
linear molecules consisting of four pyrrole rings (tetrapyrroles). All the eight pigments 
are covalently linked to the protein scaffold  by relatively stable thioether bonds involving specific Cysteines 
in the protein matrix~\cite{Scheer:2008aa}. The central pair of highly correlated PEB pigments is 
highlighted. Top: front view with the PEB50/61 pair in the front. Bottom: rear 
view with the PEB50/61 pair in the back.}
\label{f:bilins}
\end{figure}

Environmental effects can strongly influence the dynamics of quantum excitations 
in complex media. In the context of photosynthetic complexes~\cite{Amerongen:2000ly,:2014fk},
a crucial role is  played by the ultrafast process that allows to transfer the energy absorbed 
by the pigments from photons to  the reaction center during the early stages of photosynthesis.
In this respect, one of most relevant issues concerns the mechanisms that allow a fast and 
efficient energy transfer within the complex. 
The presence of a noisy protein scaffold that embeds the pigment network represents an essential
feature of the problem. Indeed, at the microscopic level chromophores experience
environment vibrations~\cite{Adolphs2006} that produce fluctuations of the local pigment site-energies
and pigment-pigment couplings. Additionally, very slow protein motions contribute to add static
disorder to the unperturbed exciton dynamics. \\
\indent  One of the key findings of recent theoretical research is that in certain regimes noise can act
as a positive factor for transport efficiency.
In particular, it has been shown~\cite{rebentrost2009environment,caruso2009,huelga2008} 
that exciton transfer efficiency can be enhanced by pure dephasing noise, that accounts for
site-energy fluctuations due to protein vibrations.
Optimal transport efficiency was found for intermediate amplitudes of the noise strength, in a region
of the parameters compatible with room temperature conditions.
The reason why a certain amount of dephasing is beneficial is that external fluctuations may promote
excitonic transitions that speed up the global transfer process. \\
\indent  Understanding the role of environmental effects on the exciton dynamics necessarily requires a microscopic 
description of the environment-induced fluctuations on the electronic degrees of freedom.
Extensive studies~\cite{Adolphs2006,rebentrost2009JPCB,fassioli2010JPCL,nalbach2010,Silbey2010NJP} 
have been performed in order to understand the role of spatio-temporal noise correlations
in site-energy fluctuations and the presence of specific vibrational modes that interact with the 
electronic ones. 
%In this respect, it is worth recalling that the usual assumption amounts to neglecting the effect of the
%environment on the pigment couplings.
Although in photosynthetic systems  site-energy fluctuations are considerably
larger than  fluctuations in the site-site coupling energies~\cite{renger2012,curutchet2013}, 
it has been shown by  Silbey and 
co-workers~\cite{Chen:2010uq,Vlaming:2012qy}
that even small, correlated coupling fluctuations may have a deep impact  for speeding up the  energy transfer process.
Moreover, it was found   that nonvanishing correlations between site-energy and coupling fluctuations may strongly
enhance or suppress long-lived oscillations in populations, depending on the sign of the correlation.
%and enhancing long-lived oscillations in populations.
These effects were initially studied within a generalized Haken-Strobl-Reineker (HSR) model~\cite{schwarzer1972moments,haken1973exactly},
where external environmental fluctuations are described  in terms of classical Gaussian Markov processes (white noise).
Further studies by Huo and Coker~\cite{huo2012influence} performed with refined numerical techniques~\cite{huo2011PLDM}
along with more recent ones by other groups~\cite{hwang2015,roden2016,Liu2016} using different modified quantum master equation
(QME) approches showed that the same phenomenology persists also with explicit non-Markovian harmonic models of heat bath.\\
\indent In this paper we  focus on the dynamics of the phycoerythrin 545 (PE545) light-harvesting 
complex of the marine cryptophyte alga Rhodomonas sp. strain
CS24, that comprises 8 bilins~\cite{curutchet2011,doust2004,doust2006,novoderezhkin2010}, see Fig.~\ref{f:bilins}.
This system absorbs very efficiently incident sunlight. Interestingly, recent studies
ascribed this efficiency to  the structural flexibility of the 
pigments, which allows the modulation of their absorption energy through local 
pigmentÐprotein interactions~\cite{curutchet2013}.
This light-harvesting system seems to work as a {\em funnel}, that drives to the surface 
the excitons formed in the core of the complex, i.e. at the two strongly coupled PEB 50/61 pigments
(see Fig.~\ref{f:bilins}). This in turn is believed to facilitate the transfer of the 
excitonic energy to another complex in the neighborhood or to the reaction center, where the 
charge separation process may occur\cite{curutchet2013}.
What is even more interesting, in a recent paper some of the present authors 
have shown that the spectral densities of individual 
pigments of PE545 may be very different with respect to 
those averaged over pseudo-symmetric pairs or over all pigments~\cite{Viani:2014kl}.
This strongly warns against an uniform parameterization of excessively simplistic quantum master equation 
approaches, such as Ref.~\cite{rebentrost2009environment},
where a single dephasing rate is assumed to describe the whole extent of possible correlations.
Rather, the need for a proper {\em ab-initio} parameterization of QME-based methods is apparent, as it is 
the whole range of possible correlations (site-energy/site-energy, coupling/coupling and site-energy/coupling)
that appears to contain the fine details of phonon-mediated exciton transport in 
light-harvesting proteins.  \\
\indent The aim of this paper is to clarify the role and the relative weight of 
the different types of correlations among 
fluctuations, involving both site energies and coupling strengths,
by parametrizing a general HSR quantum master equation directly from
atomistic Quantum-Mechanics/Molecular-Mechanics (QM/MM) simulations.
In section 2, we describe this procedure with the aim of computing  
the  transport parameters  of the complex at room
temperature and  study their dependence on the strength of the environment-induced perturbations.
Accordingly, in section 3 we show that the so-called Zeno regime~\cite{rebentrost2009environment,Misra1977zeno},
corresponding to a drop in exciton transfer efficiency at high noise strength
is suppressed by coupling fluctuations at high temperature. However, while the Zeno regime 
might be relevant for some materials, as it has been already 
been shown by some of the present authors~\cite{iubini2015NJP}, it is certainly
{\em de facto} unaccessible to biological macromolecules, 
as the protein scaffolds would fall apart well before reaching  such high temperatures.  
From a theoretical standpoint, we show that the theoretical suppression of the Zeno effect
can be simply interpreted in terms of the emergence of a  
classical random walk for the exciton dynamics. In this limit, we show that the quantum master 
equation becomes a simple classical Fokker-Planck equation, whose diffusivity tensor is given 
by the coupling-coupling correlation matrix. 
Furthermore, in section 3 we show that slow frequency components at the limits of validity of the 
HSR approach are present in specific cross-correlations. 
In section 4 we use an original elastic network model approach 
to rationalize the emergence of these low frequencies in terms of specific collective 
modes involving preferentially the subnetwork of cysteines that are covalently linked 
to the pigments. In section 5 we epitomize our results, discussing critically  
HSR based approaches and suggesting an alternative way to dissect the fold-rooted spatiotemporal 
signatures of exciton transport in light-harvesting structures based on elastic network models.

%=================================================================================================================
%=================================================================================================================

%=================================================================================================================
\section{Reduced quantum master equation from QM/MM simulations}

\begin{figure}%[t!,clip]
\begin{center}
\includegraphics[width=0.5\textwidth]{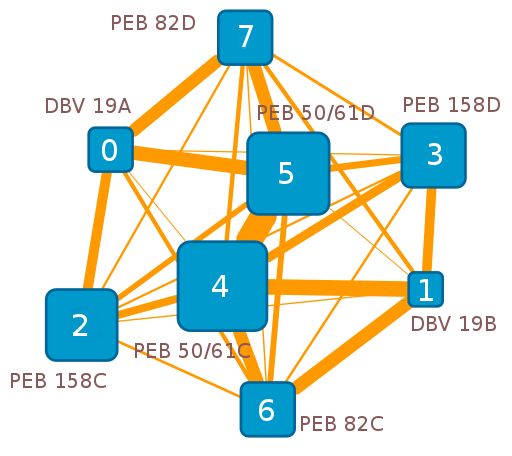}
\end{center}
\caption{The exciton network corresponding to the Hamiltonian $H$ of the PE545 light-harvesting 
antenna according to the parameters reported in Ref.~\cite{curutchet2013}. The sizes of nodes
and edges account for the amplitude of average pigment site-energies and couplings, respectively.
The spatial arrangement of nodes reflects the  position of chromophores in the complex, 
see Fig.~\ref{f:bilins}, rear view. In particular, the two central PEBs (50/61) appear clearly 
to form the most {\em connected} pair in the system.}.
\label{f:network}
\end{figure}

The excitonic two-level system consisting of $N=8$ pigments of the light-harvesting
complex can be described by the tight-binding Hamiltonian given by
\begin{equation}
\label{tight-binding}
H_0= \sum_{m=0}^{N-1} \epsilon_m |m \rangle \langle m| + \sum_{n<m}^{N-1} J_{mn} \Big(|m \rangle \langle n| +
|n \rangle \langle m|\Big)
\end{equation}
where the state $ |m \rangle $ denotes an electronic excitation localized on chromophore $m$. 
The site energies and pigment-pigment couplings are given by $ \epsilon_m $ and $ J_{mn} $ respectively.
For the parametrization of the Hamiltonian $H_0$ of the bilin system we refer to~\cite{curutchet2013}, 
see also the online supporting information. 
The associated network is shown in Fig.~\ref{f:network}, where we identify the nodes of the network  from $m=0$
to $m=N-1=7$ with the eight bilin molecules of the complex and (weighted) 
links represent the associated coupling strengths.
%
%%%%%%%%%%%%%%%%%%%%%%%%%%%%%%%%%%%%%%%%%%%%%%%%%%%%%%%%%%%%%%%%%%%%%%%%%%%%%%%%%%%%%%%%%%%%%%%%%%%%%%%%%%%%%%%%%%%
%%%%%%%%%%%%%%%%%%%%%%%%%%%%%%%%%%%%%%%%%%%%%%%%%%%%%%%%%%%%%%%%%%%%%%%%%%%%%%%%%%%%%%%%%%%%%%%%%%%%%%%%%%%%%%%%%%%
% 
To include the effect of a thermally fluctuating  environment, the original approach of Haken and 
Strobl~\cite{schwarzer1972moments,haken1973exactly}  requires one to solve a modified problem, 
described by the Hamiltonian
\be
\label{eq:exct_full_H}
H(t) = H_0 + h(t)
\ee
where the Hermitian operator $h(t)$ is a time-dependent perturbation of the Hamiltonian $H_0$,
whose temporal evolution reflects the bath dynamics.
In this framework, the so called Haken-Strobl-Reineker  model~\cite{haken1973exactly,Vlaming:2012qy} assumes  
that the environmental fluctuations $h(t)$ are classical Gaussian 
Markovian processes with zero average and correlations given by
\be\label{eq:markovian}
C_{mnpq}(t-t')\equiv
\left\langle h_{mn}(t)h_{pq}(t') \right\rangle = 2 \delta(t-t') \Gamma_{mnpq}
\ee
where  the four-indices tensor $\Gamma_{mnpq}$ is invariant under the following set of 
index permutations~\cite{Vlaming:2012qy}
\be\label{eq:symm}
(mn) \rightarrow (nm) \quad (pq) \rightarrow (qp) \quad (mnpq) \rightarrow (pqmn)
\ee
With the above hypotheses, the information on the external environment is fully encoded in the 
tensor $\Gamma_{mnpq}$.\\
\indent Finally, to account for the exciton losses due to internal recombination processes 
and exciton transfer to the reaction center, we add to the system Hamiltonian $H_0$ an 
anti-Hermitian operator defined as
\be\label{eq:hdiss} 
iH_d = - i \hbar  \sum_{m} (\alpha_m + \kappa_m) |m \rangle \langle m| 
\ee
In the above formula, $ \alpha_m $ and $ \kappa_m $ are the site-dependent 
recombination and trapping rates, respectively.
Altogether, it can be shown~\cite{haken1973exactly} that the evolution of the density matrix of the system is
described by the following master equation
\be
\begin{split}\label{eq:Silbey}
\frac{d \rho_{nm}}{dt}=& -i\left[ H_0,\rho \right]_{nm} -i\left[ H_d,\rho \right]^+_{nm}  \\
                      +&\sum_{pq}\left( 2\Gamma_{npqm} \rho_{pq}
- \Gamma_{nppq}\rho_{qm} -\Gamma_{pmqp}\rho_{nq} \right)
\end{split}
\ee
where the symbols $\left[ \cdot,\cdot \right]$ and $\left[ \cdot,\cdot \right]^+$ denote the commutator and
the anti-commutator respectively.
The first term in the r.h.s of  Eq.~(\ref{eq:Silbey}) describes the unitary part of 
the evolution, the second one incorporates
the dissipative effects defined in Eq.~(\ref{eq:hdiss}), while the third one accounts for the action of the 
incoherent noise.\\
\indent Following Vlaming and Silbey~\cite{Vlaming:2012qy}, it is instructive to single out three physically 
different types of elements of the tensor $\Gamma$, namely:
\begin{enumerate}%[I]
 \item Site energy-site energy correlations, $\Gamma_{mmnn}$
 \item Site energy-coupling correlations, $\Gamma_{mmpq}$ with $p\neq q$
 \item Coupling-coupling  correlations, $\Gamma_{nmpq}$ with $n\neq m$ and $p \neq q$
\end{enumerate}
In the following subsection we will discuss how to estimate the bath 
tensor $\Gamma$ directly from atomic-level QM/MM simulations.\\
%
%-----------------------------------------------------------------------------------------------------------------------------------
\subsection{Microscopic description of the exciton environment}

All the excitonic parameters (site energies and excitonic couplings in $H(t)$, see 
Eq.~(\ref{eq:exct_full_H})) have been calculated
with a polarizable quantum mechanical/molecular mechanical (QM/MMPol) 
methodology~\cite{curutchet2009electronic} using a trajectory obtained with a ground-state 
classical MD simulation of PE545 in water (see Ref.~\cite{curutchet2011} for more details). 
In particular, we extracted 60000 snapshots every 5 fs 
corresponding to 300 ps of a MD simulation of PE545 for the subsequent QM/MMPol calculations. 
Because of the considerable computational cost involved in the QM/MMPol calculations, 
we adopted the ZINDO semi-empirical method~\cite{zerner1991semiempirical}  to describe the excited states of the pigments,
whereas the protein and the solvent were described using a polarizable force field. 
All details about the QM/MMPol model are reported in Refs~\cite{curutchet2013} and~\cite{curutchet2011}.
A detailed study on the parametrization of $H_0$ for PE545  is contained in Ref.~\cite{curutchet2013}. 
Concerning the environment fluctuations described by the term $h(t)$,
the white noise approximation of the HSR model, Eq.~(\ref{eq:markovian}), assumes that the bath degrees
of freedom decorrelate much faster than the typical exciton timescales. 
Therefore, in this limit, one can formally integrate Eq.~(\ref{eq:markovian}), writing
\be\label{eq:int_markovian}
2\Gamma_{mnpq} = \int_{-\infty}^{+\infty} C_{mnpq}(\tau) \, d\tau
\ee
where it clearly appears that $\Gamma_{mnpq}$ are related to the zero-frequency 
limit of the Fourier transform  $C_{mnpq}(\omega)$.
Notice that, despite the integration up to arbitrarily large times in the above formula, the relevant contribution
of correlations to the tensor $\Gamma$ is expected to come exclusively from very short times.
An instructive example is offered by the case of an exponential decay of the correlation function,
$C(\tau) = C(0) \exp(-|\tau|/\tau_c )$, which gives $\Gamma=C(0)\tau_c$. In this example $\tau_c$ defines 
the characteristic timescale of bath correlations, so that the contribution to the integral in 
Eq.~(\ref{eq:int_markovian}) for  times $|\tau|\gg\tau_c$ is exponentially small in $\tau$. \\
\indent By means of  Eq.~(\ref{eq:int_markovian}), we have computed the elements of the tensor $\Gamma_{mnpq}$ from the
environment correlation functions $C_{mnpq}$ extracted from QM/MM trajectories.
An effective infrared threshold $\Theta$ was  imposed, in order  to exclude the contribution of slow environmental 
correlations which are inconsistent with the HSR white-noise approximation of the exciton-protein coupling dynamics.
Accordingly,  in Eq.~(\ref{eq:int_markovian}) we have neglected the 
correlations displaying a decaying time $\tau_c > \Theta$
(see the supplemental material for further details).\\
\indent The choice of $\Theta$ is a rather delicate point, since on the one hand
$\Theta$ should be taken as small as possible for  the Markovian approximation to hold, on the other
the limit $\Theta=0$ yields vanishing couplings $\Gamma_{mnpq}$, thus washing out the whole  
environment in the HSR description.
A detailed study  of the optimal choice of  $\Theta$ for the description of the PE545 protein environment
is beyond the scope of the present paper. 
Here, we take a pragmatic approach by focusing on the upper bound 
of $\Theta$, that is set by the intrinsic excitonic timescales.
In particular, we have chosen $\Theta=1$ ps, which is comparable to the typical exciton transfer timescales 
found in Ref.~\cite{curutchet2013} for the PE545 complex.  An analysis of the role of $\Theta$ for the amplitude
of the HS coefficients  is provided at the end of Section~\ref{sec:res_gamma}. 

%================================================================================================================================
\section{HSR coefficients: the white noise approximation}\label{sec:res_gamma}

\begin{figure}%[t!,clip]
\begin{center}
\includegraphics[width=0.7\textwidth]{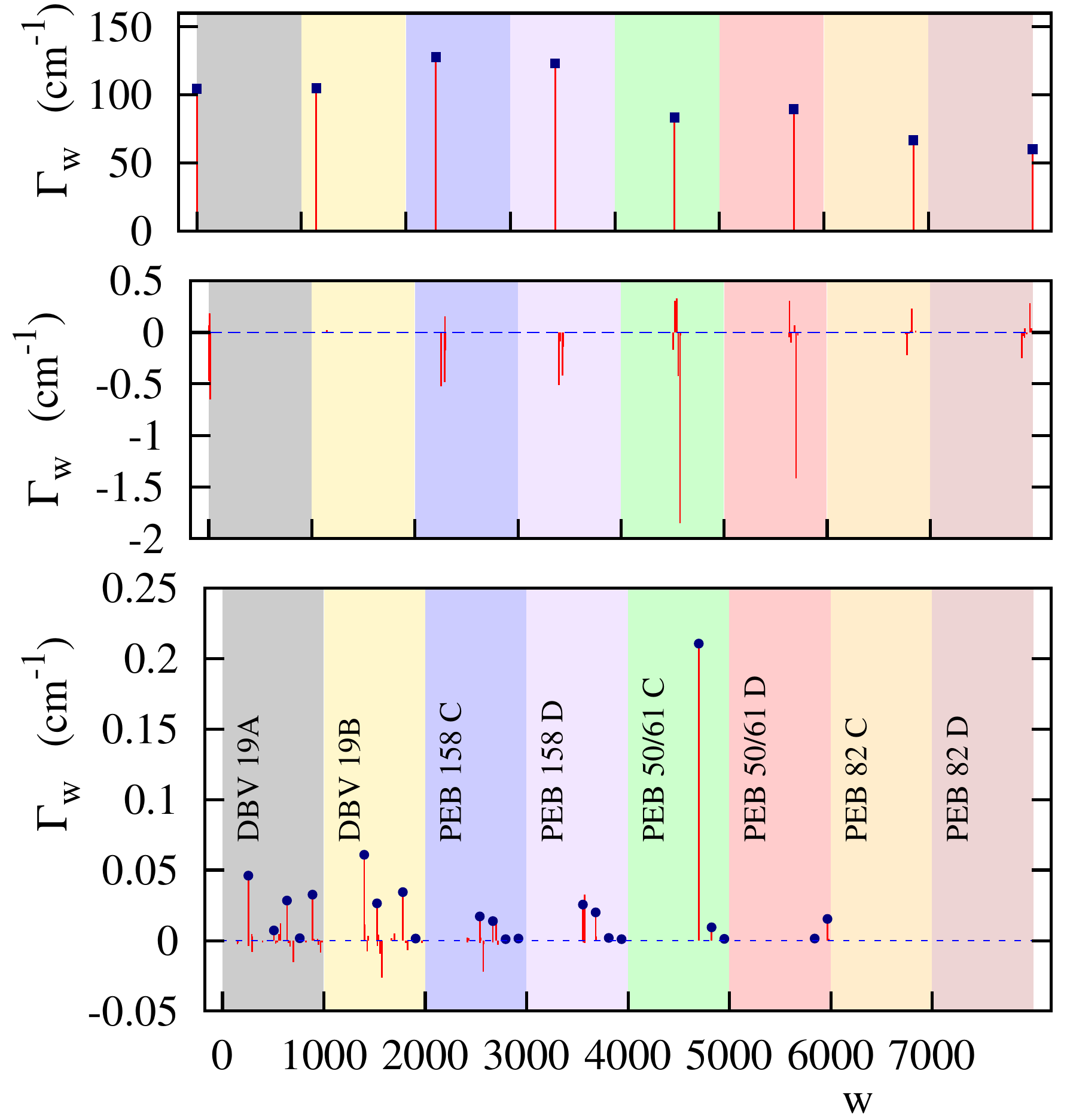}
\end{center}
\caption{Bath parameter $\Gamma_w$ extracted from QM/MM simulations at $T=300K$. 
The index $w$ is the octal representation of the number $mnpq$.
Upper panel: site energy autocorrelation terms of type $\Gamma_{mmmm}$ (sector $i$);
central panel: site energy-couplings correlations of type $\Gamma_{mmpq}$ with $p\neq q$ (sector $ii$);  
lower panel: Coupling-coupling terms of type $\Gamma_{mnpq}$, with $m<n$, $p<q$, $m\leq p$, $n\leq q$ (sector $iii$).
Blue dots represent coupling autocorrelations, i.e. terms $\Gamma_{mnmn}$. 
Each colored stripe delimits all the possible combinations of the kind $mnpq$ for each pigment index $m$. 
}
\label{f:Gamma_tensor}
\end{figure}

\begin{figure}%[t!,clip]
\begin{center}
\includegraphics[width=0.8\textwidth,clip]{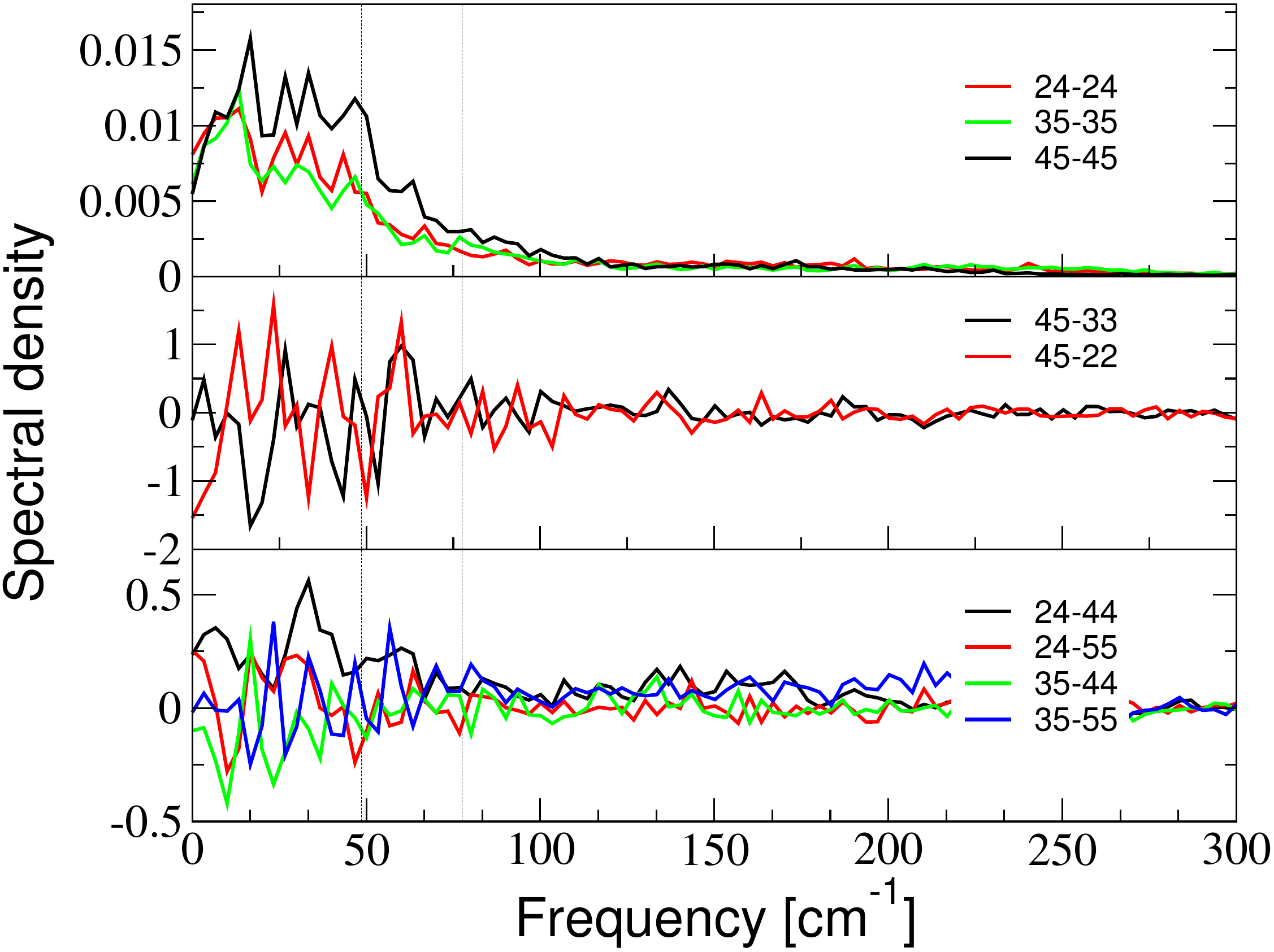}
\end{center}
\caption{
Cosine Fourier transform of selected auto and cross correlation functions. For 
example, the black and red lines in the middle panel refer to the cross correlations between 
the fluctuations of the PEB50/61C-PEB50/61D coupling (45) and the PEB158D (3) and PEB158C (2)
site energies, respectively.
The numbering of pigments is the one illustrated in Fig.~\ref{f:network}. The vertical dashed lines delimit the frequency 
region where our ENM analysis identify normal modes expressing concerted motions of the 
sub-network of CYS residues covalently linked to the pigments through disulfide bridges (see text).   
The spectra in the upper panel have been normalized to unit area for the sake of clarity. 
The others have units of cm.
}
\label{f:sel_spectr}
\end{figure}

An overall view of the relevant Haken-Strobl couplings obtained with this method provided in Fig.~\ref{f:Gamma_tensor},
where we separately plot the elements of kind i, ii and iii up to the permutations stated in Eq.~(\ref{eq:symm}). 
Here, in order to represent unambiguously the elements $\Gamma_{mnpq}$, we have introduced an octal integer $w$  
ranging from 0 to 7777, whose digits are $mnpq$. For each sector, we therefore plot the nonvanishing coefficients $\Gamma_w$. 
%
%%%%%%%%%%%%%%%%%%%%%%%%%%%%%%%%%%%%%%%%%%%%%%%%%%%%%%%%%%%%%%%%%%%%%%%%%%%%%%%%%%%%%%%%%%%%%%%%%%%%%%%%%%%%%%%%%%%%%%%%%%%%%%%%%%
%%%%%%%%%%%%%%%%%%%%%%%%%%%%%%%%%%%%%%%%%%%%%%%%%%%%%%%%%%%%%%%%%%%%%%%%%%%%%%%%%%%%%%%%%%%%%%%%%%%%%%%%%%%%%%%%%%%%%%%%%%%%%%%%%%
%
Consistently with other studies~\cite{renger2012}, we find that the relevant contribution 
(order $100$ cm$^{-1}$) in the tensor $\Gamma_w$ is given by the diagonal terms $\gamma_{mmmm}$ 
in sector $i$. Notice however that these terms display a 
clear dependence on the pigment site $m$ with an approximate correlation induced by the global geometry of the complex, as also 
emphasized in Ref.~\cite{Viani:2014kl} --
see for instance the two lowest terms
$\Gamma_{6666}$ and $\Gamma_{7777}$, which correspond to ``antipodal'' pigments 
PEB 82C and PEB 82D in Fig.~\ref{f:bilins}.
Still within sector $i$, we have found no relevant contributions from site-energies
cross correlations,  i.e. $\Gamma_{mmnn}$ with $m\neq n$. This is consistent with a previous 
study by Schulten et {\it al.} for the FMO complex~\cite{schulten2011} and reinforces 
the idea that in general fluctuations in the excitation energies on different
pigments are  not significantly  correlated on the time scales of exciton transfer. 
Concerning sector $ii$ correlations, the maximum amplitude of $\Gamma_w$ is of
order $1$ cm$^{-1}$ in absolute value. Moreover, we find that large couplings typically
correspond to anticorrelations (i.e. negative $\Gamma_w$)
between the site energy of a pigment $m$ and a coupling involving \emph{the same} pigment $m$,
i.e. for $w=mmmn$ or $w=mmnm$ with $m\neq n$.
This last property was already spotlighted in Ref.~\cite{viani2013spatial} 
from the analysis of static correlation functions.
A more detailed presentation of the environment couplings in sector $ii$ is provided in the 
supplemental material.\\
%
%
%%%%%%%%%%%%%%%%%%%%%%%%%%%%%%%%%%%%%%%%%%%%%%%%%%%%%%%%%%%%%%%%%%%%%%%%%%%%%%%%%%%%%%%%%%%%%%%%%%%%%%%%%%%%%%%%%%%%%%%%%%%%%%%%%%
\begin{figure}%[t!,clip]
\begin{center}
\includegraphics[width=0.7\textwidth,clip]{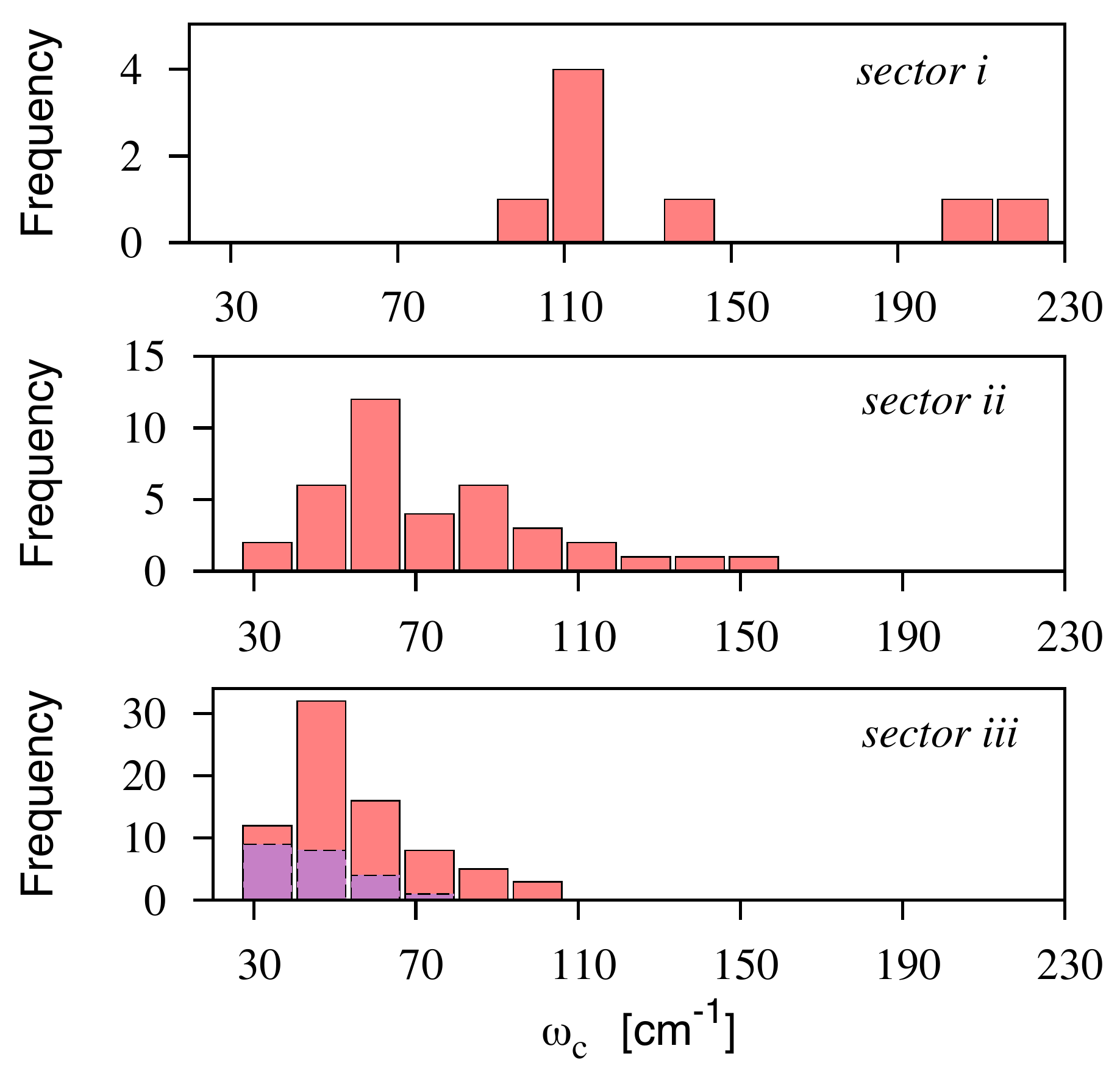}
\end{center}
\caption{Histograms of the spectral widths $\omega_c$  extracted by fitting a Lorentzian line shape
to the cosine Fourier transforms of the autocorrelation functions $C_{mnpq}(\omega)$ in the range $[0,500]$ cm$^{-1}$.   
The three panels show the distribution of $\omega_c$ in each correlation 
sector. For sector $iii$  the purple bars refer to coupling-coupling autocorrelations,
i.e. to $C_{mnmn}(\omega)$ with $m \neq n$. Histograms are computed for
$\omega_c \geq 2\pi/\Theta \simeq 33$ cm$^{-1}$.
}
\label{f:distr}
\end{figure}
%%%%%%%%%%%%%%%%%%%%%%%%%%%%%%%%%%%%%%%%%%%%%%%%%%%%%%%%%%%%%%%%%%%%%%%%%%%%%%%%%%%%%%%%%%%%%%%%%%%%%%%%%%%%%%%%%%%%%%%%%%%%%%%%%%
%
\indent Here, we observe that the largest couplings that emerge in this sector are
$\Gamma_{4445}=-1.8$ cm$^{-1}$ and $\Gamma_{5545}=-1.4$ cm$^{-1}$ (see the  central panel of Fig.~\ref{f:Gamma_tensor}).
These correspond to the correlations between the site energies and the coupling of the strongly-interacting central
pair of pigments PEB 50/61C and PEB 50/61D, see Fig.~\ref{f:network}.
It is therefore not surprising that all the other couplings in sector $ii$ are significantly smaller than these two ones.
Moreover, according to the analytical results found by Chen and Silbey~\cite{Chen:2010uq} in the HSR model, environment
couplings in sector $ii$ are expected to trigger population oscillations depending on the difference  $\Gamma_{mmmn}-\Gamma_{nnmn}$.
Since  the relevant terms $\Gamma_{4445}$ and $\Gamma_{5545}$   have the same sign and comparable absolute amplitude, 
we do not expect any relevant phenomenon related to long-lived oscillations in the HSR approximation.\\
\indent Moving finally to sector $iii$, Fig.~\ref{f:Gamma_tensor} shows that the 
amplitude of all $\Gamma_w$ falls below $0.1$ cm$^{-1}$ except for the 
term $\Gamma_{4545}$ which is representative of the coupling autocorrelations of the central
couple of pigments. More generally, we find that the non-vanishing  terms in this sector typically
correspond to coupling autocorrelations, in the form $\Gamma_{mnmn}$ with $m\neq n$, 
(see the blue dots in Fig.~\ref{f:Gamma_tensor}).
On the other hand, coupling cross correlations $\Gamma_{mnpq}$ with $m \neq n$, $p \neq q$ and $mn \neq pq$
provide   a marginal contribution  to  sector $iii$, typically of the order 
of $10^{-3}$ cm$^{-1}$.\\
\indent A complementary view on the extracted coefficients $\Gamma_{mnpq}$ can be obtained by
formally estimating a time scale for the decay of the associated time correlations. In the simplest 
hypothesis of a single finite time decay constant, one should replace  eq.~(\ref{eq:markovian})
by 
\be\label{eq:markovian_exp}
C_{mnpq}(t-t')\equiv
\left\langle h_{mn}(t)h_{pq}(t') \right\rangle = \frac{2 e^{-|t-t'|/\tau_c}}{\tau_c}  \Gamma_{mnpq}
\ee
Accordingly an estimate of the spectral width $\omega_c=2\pi/\tau_c$ can be obtained 
from the QM/MM data by consistently fitting a Lorentzian to the cosine Fourier transform 
of the correlation functions.
Fig.~\ref{f:distr}  displays the result of this analysis for each sector.
While site energy autocorrelations (sector $i$) display a fast decay, correlations in sectors $ii$ and 
$iii$ decay with timescales that are comparable with the typical exciton timescales~\cite{curutchet2013}.
More precisely, we find the following ordering with respect to average decreasing $\omega_c$:
site-energy autocorrelations, site-energy-coupling correlations, couplings cross correlations, 
coupling autocorrelations. This analysis proves that coupling-coupling autocorrelations lie at the 
limits of validity of the white-noise approximation intrinsic to the HSR approach. In this sense, 
exciton transport in proteins would possibly require more sophisticated methods such as 
used in organics semiconductors, where it is known that coupling-coupling fluctuations are 
dominated by low-frequency, collective vibrations of the molecular 
crystal~\cite{Troisi2011,Troisi:2006fk,Arago:2015bh}
More precisely, in the case of the PE545 protein it turns out that signatures of slow 
frequency components emerge in the correlations among selected 
coupling and/or site energy fluctuations. A few typical examples illustrating this behaviour are 
reported in Fig.~\ref{f:sel_spectr}. Fluctuations with frequencies of the order of a few tens 
of cm$^{-1}$ are likely to flag low-frequency, collective normal modes (NM) of the protein matrix, which 
could potentially spotlight important, fold-rooted motions expressed by the protein scaffold with 
the aim of selectively promoting  transport by helping break localization of exciton wave-functions
induced by disorder in the site energies~\cite{Plenio:2008fk,rebentrost2009environment}.
In section~\ref{sec:res_NM} we shall dig further into this hypothesis, with the aim of uncovering whether 
the PE545 protein structure is indeed able to express specific, potentially functional, transport-enhancing
normal modes in the highlighted frequency range.  In the next section, we stick to the 
white-noise approximation and turn to investigate the effect of the QM/MM-based correlations on 
exciton transport.

%%%%%%%%%%%%%%%%%%%%%%%%%%%%%%%%%%%%%%%%%%%%%%%%%%%%%%%%%%%%%%%%%%%%%%%%%%%%%%%%%%%%%%%%%%%%%%%%%%%%%%%%%%%%%%%%%%%%%%%%%%%
%%%%%%%%%%%%%%%%%%%%%%%%%%%%%%%%%%%%%%%%%%%%%%%%%%%%%%%%%%%%%%%%%%%%%%%%%%%%%%%%%%%%%%%%%%%%%%%%%%%%%%%%%%%%%%%%%%%%%%%%%%%

%==========================================================================================================================
\subsection{Exciton transfer efficiency}
%==========================================================================================================================

\begin{figure}%[t!,clip]
\begin{center}
\includegraphics[width=0.8\textwidth]{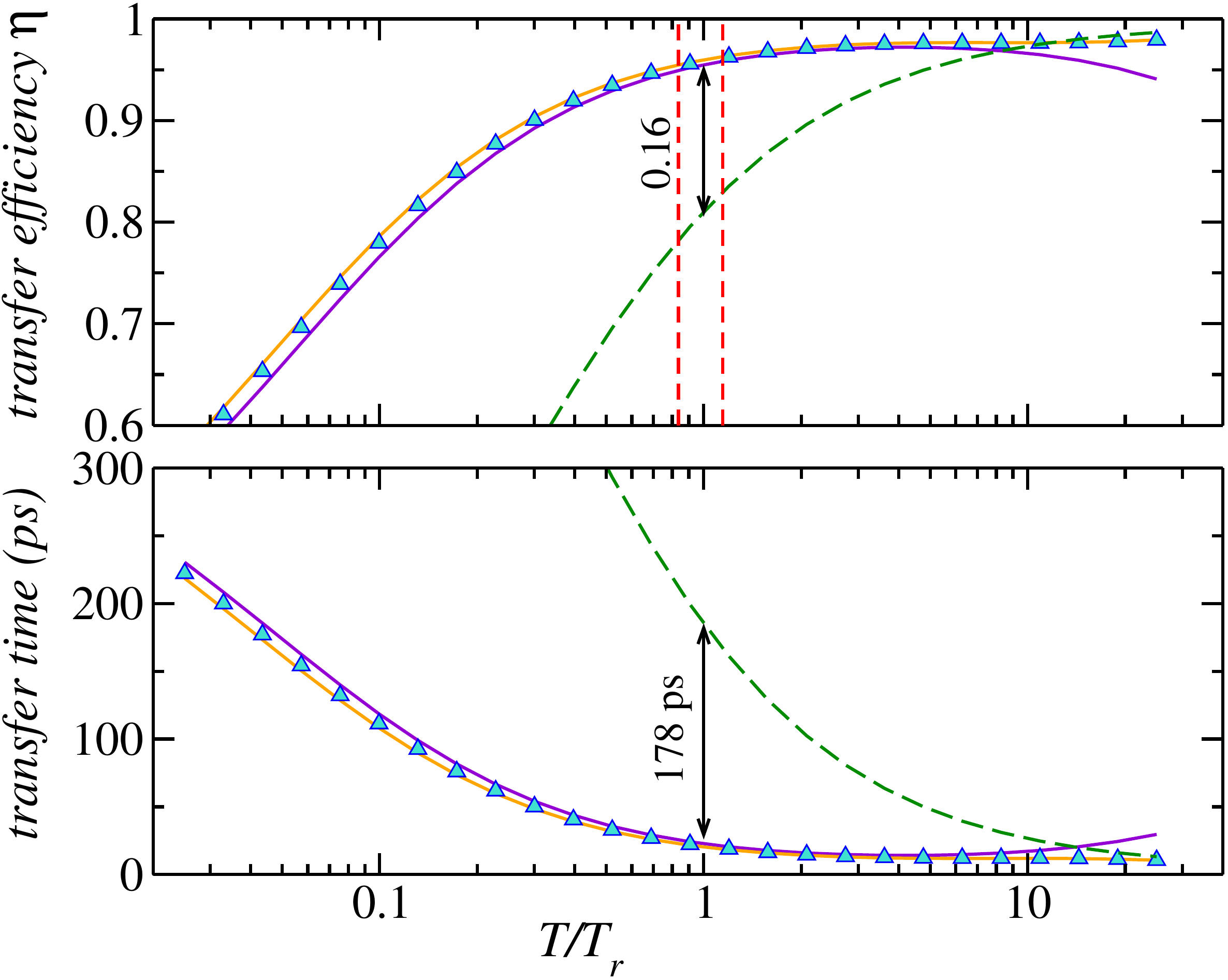}
\end{center}
\caption{Exciton transfer efficiency (top) and average transfer time (bottom) 
as a function of temperature for the exciton Hamiltonian $H_0$.
The purple curve refers to the evolution where only site energy fluctuations are allowed 
(pure dephasing noise).  The orange curve accounts for the dynamics in Eq.~\ref{eq:Silbey} with the 
full tensor $\Gamma_w$ shown in Fig.~\ref{f:Gamma_tensor}, while cyan triangles are obtained considering 
only site energy autocorrelations and coupling autocorrelations. The green dashed line shows  the result 
obtained from a purely incoherent population transfer, see Eq.~(\ref{eq:classical}).
The arrows identify the differences between the quantum and classical (purely incoherent) models at  
room temperature, while the vertical dashed lines in the top panel identify the typical stability region of 
proteins, $(240 \pm 17 < T < 343 \pm 14)$ K~\cite{Sandeep-Kumar:2003aa}.
Simulations refer to dissipation rates $\alpha_m=\alpha=1$ ns$^{-1}$ and
$\kappa_m=\kappa\,\delta_{1,m}$ with $\kappa=1$ ps$^{-1}$. 
}
\label{f:efficiency}
\end{figure}

The exciton transfer process can be quantified by means of two simple indicators, namely the transfer
efficiency $\eta$ and the average transfer time $\bar t$~\cite{rebentrost2009environment}. These 
quantities can be computed simply in terms of time evolution of the one-exciton density matrix,
\begin{eqnarray}
\eta &=& 2 \sum_{m}  \kappa_m \int_{0}^{\infty} dt \langle m| \rho(t) |m\rangle \label{eq:efficiency}\\
\bar t &=& \frac{2}{\eta} \sum_{m}  \kappa_m \int_{0}^{\infty} dt \,t\, \langle m| \rho(t) |m\rangle \label{eq:avtime}
\end{eqnarray}
The presence of a finite recombination rate $\alpha_m$ in Eq.~(\ref{eq:hdiss}) introduces a non-specific channel 
for exciton losses, that reduces the energy flux towards the specific trapping site. Therefore, the efficiency parameter
$\eta\in[0,1]$ quantifies the  probability that an exciton is successfully driven to the trap before it recombines.
A typical value of $\alpha_m$ is estimated to be independent of $m$ and of order $1$ ns$^{-1}$~\cite{rebentrost2009environment}.
On the other hand, we assume the presence of a single trapping site located on the lowest-energy pigment ($m=1$)
($DBV_{19B}$)~\cite{curutchet2013} with  trapping rate of $1$ ps$^{-1}$. 
In order to explore the role of environment fluctuations in Eq.~\ref{eq:Silbey}, we introduce 
a linear dependence of $\Gamma_w$ on the system temperature $T$, namely $\Gamma_w(T) = (T/T_r) \Gamma_r$ where $T_r=300$ K 
denotes room temperature and $\Gamma_r$ is the coupling tensor at $T=T_r$ computed from QM/MM simulations 
as discussed in Section~\ref{sec:res_gamma}.\\
\indent In Fig.~\ref{f:efficiency} we show the parameters $\eta$ and $\bar t$ as a function of temperature 
for an exciton initially localized on pigment  PEB50/61D (site $m=5$), i.e. one of the  bilins forming the 
central pair of strongly interacting pigments (see Fig.~\ref{f:network}).
Similarly to previous studies on the FMO complex~\cite{rebentrost2009environment}, we find that
at room temperature $T=T_r$, the PE545 complex has almost reached the condition of maximum 
efficiency and minimum transfer time. Moreover, this quasi-optimum condition is by all practical means
rather insensitive to temperature (note the logarithmic scale on the $x$-axis) in a range of temperatures 
where there can be safely asserted that the protein is in its folded, fully functional structure. 
It is worthwhile to observe that this temperature range is rather narrow, typically $\pm 50$ K 
around room temperature. It is evident from Fig.~\ref{f:efficiency} that the efficiency and average transfer 
times are rather constant in this temperature range (see vertical dashed lines in the top panel). 
It should be reminded that, strictly speaking, it is meaningless to consider exciton transport in light-harvesting 
systems outside this temperature range. \\
%
%%%%%%%%%%%%%%%%%%%%%%%%%%%%%%%%%%%%%%%%%%%%%%%%%%%%%%%%%%%%%%%%%%%%%%%%%%%%%%%%%%%%%%%%%%%%%%%%%%%%%%%%%%%%%%%%%%%%%%%%%%%%%%%
%%%%%%%%%%%%%%%%%%%%%%%%%%%%%%%%%%%%%%%%%%%%%%%%%%%%%%%%%%%%%%%%%%%%%%%%%%%%%%%%%%%%%%%%%%%%%%%%%%%%%%%%%%%%%%%%%%%%%%%%%%%%%%%
%
\indent In order to gain a deeper insight on the overall transport phenomenon,
we have repeated the same numerical experiment by  selectively ``silencing'' parts of
the dynamical contributions appearing in Eq.~\ref{eq:Silbey}.  
A first important result is that, for the wide range of temperatures here considered (biologically 
relevant and irrelevant), negligible variations
are observed  upon neglecting all the possible cross-correlations in $\Gamma_w$ (see cyan triangles in 
Fig.~\ref{f:efficiency}). This suggests that cross correlations in sectors $ii$ and $iii$ do not play a 
relevant role for the transport properties of the bilin complex. A complete definition of HSR coupling parameters 
in the limit of vanishing cross correlations is provided  in the supplemental material.
Further insight can be gathered from a further reduction of $\Gamma_w$ to a 
diagonal tensor with non-vanishing elements of kind $\Gamma_{mmmm}$, i.e.
the coefficients  shown in the upper panel of Fig.~\ref{f:Gamma_tensor}. The above approximations lead to
the usual description of environmental effects in terms of {\em pure dephasing} noise, 
where optimal transport was shown to emerge~\cite{rebentrost2009environment} for a {\it finite} dephasing rate.
It is clear from Fig.~\ref{f:efficiency} that the characteristic high-temperature efficiency drop (associated
in Ref.~\cite{rebentrost2009environment} with the quantum Zeno effect~\cite{Misra1977zeno})
is no longer present when  generalized correlations beyond pure dephasing are considered. 
This result is a direct consequence of the presence of small but  non-vanishing coupling 
fluctuations, i.e. of terms $\Gamma_{mnmn}$ with $(m\neq n)$, which promote a  population transfer between sites 
$m$ and $n$~\cite{schwarzer1972moments,haken1973exactly,Chen:2010uq,Vlaming:2012qy}.
A similar effect  was recently observed in Ref.~\cite{iubini2015NJP,moix2013} within the 
framework of an explicit nonlinear quantum-classical hybrid model.
As a general observation, we emphasize that the Zeno-like decrease in efficiency associated with pure 
dephasing falls in a temperature range where the protein matrix would be in a wildly denatured state, which means 
that such regime is by all means not biologically relevant and could not have been possibly accessed 
by evolution. We observe that this conclusion would hold {\em a fortiori} were the mapping between 
temperature and strength of correlations nonlinear, i.e.  $\Gamma_w(T) = (T/T_r)^\nu \Gamma_r$ 
with $\nu > 1$.\\
\indent The high-temperature regime can be captured by a simple master equation for purely incoherent population 
transfer (a classical random walk) in the form
\be
\label{eq:classical}
\frac{d \rho_{nn}}{dt} = \sum_{m\neq n}\Gamma_{nmnm}\rho_{mm} - \rho_{nn}\sum_{m\neq n}\Gamma_{nmnm}  \quad,
\ee
where only coupling autocorrelations are considered (see green dashed line in Fig.~\ref{f:efficiency}).
Eq.~\eqref{eq:classical} is essentially a Fokker-Planck equation for the classical occupation
probability $\rho_{nn}$ on pigment $n$. As shown in Fig.~\ref{f:efficiency}, the purely classical description 
approaches the generalized HSR dynamics in the region $T\gtrsim 10\,T_r$.  It is worth noting that 
the purely diffusive description underestimates the transfer efficiency by about 17 \% (and correspondingly overestimates 
the average transfer time of a larger factor of about 8) 
in the biologically relevant temperature region (see arrows 
in Fig.~\ref{f:efficiency}), suggesting that 
some degree of coherence is indeed important to achieve optimal transport.\\
%
%%%%%%%%%%%%%%%%%%%%%%%%%%%%%%%%%%%%%%%%%%%%%%%%%%%%%%%%%%%%%%%%%%%%%%%%%%%%%%%%%%%%%%%%%%%%%%%%%%%%%%%%%%%%%%%%%%%%%%%%%%%%%%%
\begin{figure}%[t!,clip]
\begin{center}
\includegraphics[width=0.8\textwidth]{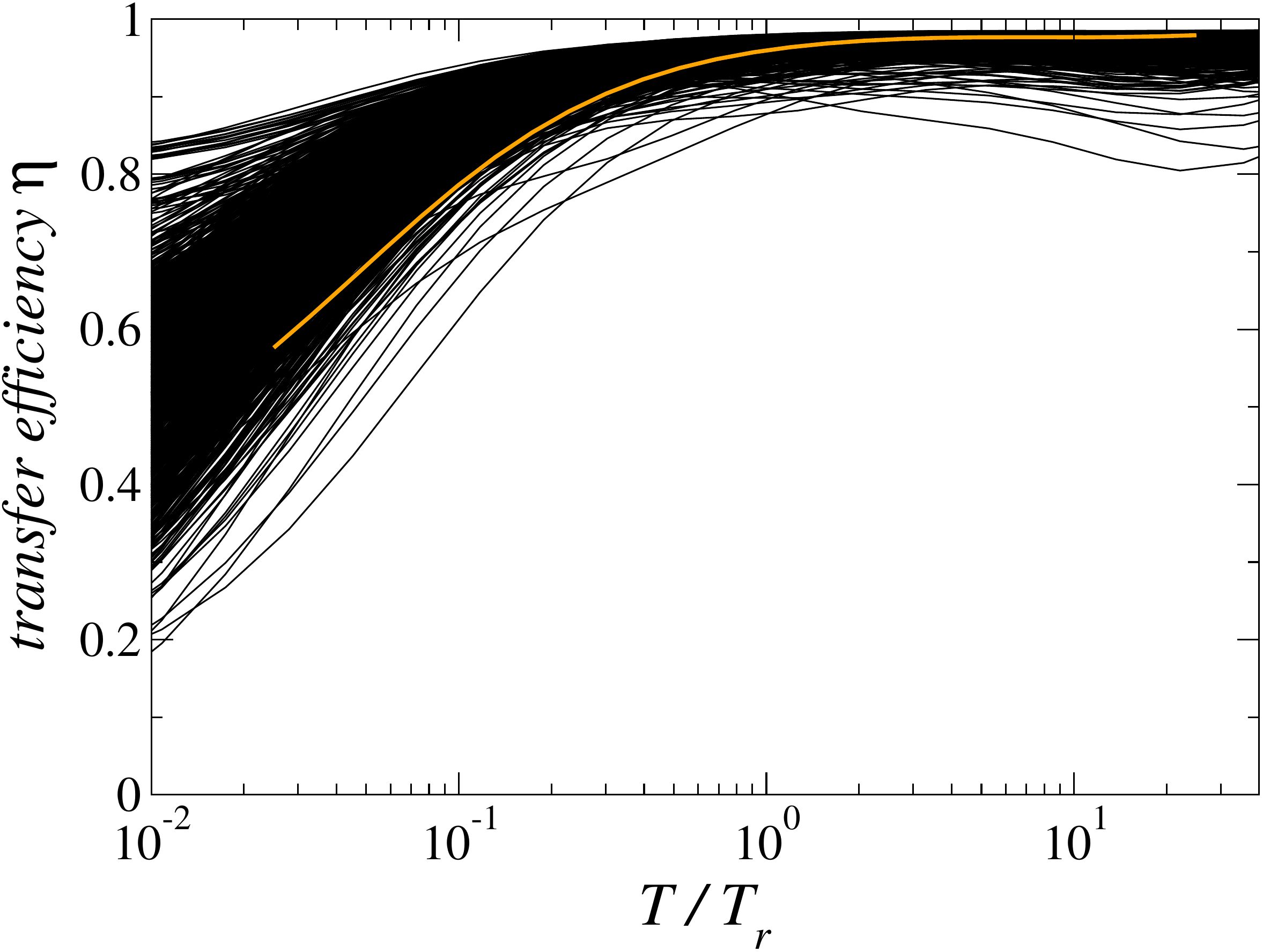}
\end{center}
\caption{Exciton transfer efficiency as a function of temperature for different modified Hamiltonians obtained 
by randomly reshuffling the off-diagonal elements $J_{nm}$ of $H_0$ (see Eq.~\eqref{tight-binding}) 
and the corresponding QM/MM fluctuations (corresponding to coupling-coupling autocorrelations). 
Simulations refer to 1000 independent reshufflings. Other parameters were set as in Fig.~\ref{f:efficiency}.
The orange line denotes the results shown in Fig.~\ref{f:efficiency} obtained with the 
full tensor $\Gamma_w$ shown in Fig.~\ref{f:Gamma_tensor}.
}
\label{f:effreshuffled}
\end{figure}
%%%%%%%%%%%%%%%%%%%%%%%%%%%%%%%%%%%%%%%%%%%%%%%%%%%%%%%%%%%%%%%%%%%%%%%%%%%%%%%%%%%%%%%%%%%%%%%%%%%%%%%%%%%%%%%%%%%%%%%%%%%%%%%
%
%  Reshuffling analysis
%
\indent 
From the previous analyses it is clear that a certain amount of noise is required 
to break disorder-induced localization of the exciton caused by the heterogeneity 
of local site energies. 
An interesting question to ask is what is the role of the specific {\em realization} of such noise, 
in terms of the correspondence between the elements of the tensor $\Gamma_{mnpq}$ and 
specific pairs of pigments in the structure. One simple way to address this question is to 
investigate the effect on transport of randomly reshuffling the the tensor elements.
For this purpose, one can restrict to site-energy and coupling-coupling 
auto-correlations (it is clear from Fig.~\ref{f:efficiency} that this will not induce appreciable 
differences on the results). 
In terms of the associated tight-binding network (see Fig.~\ref{f:network}), this amounts to 
randomly rewiring the nodes. Fig.~\ref{f:effreshuffled} shows efficiency curves computed for a large 
number of independent random rewiring realizations. Each of this corresponds to a different random 
reshuffling of the pairs $(J_{nm},\Gamma_{nmnm})$, with $n\neq m$,  i.e. 
pigment couplings and associated pigment-pigment noise.  It is clear that in the physically relevant 
region close to $\Gamma_r$, deviations from the efficiency corresponding to the 
{\em true} set of physical parameters are negligible.  Analogous results were obtained when considering 
random reshufflings of site energies and associated noise terms, i.e. the pairs $(J_{mm},\Gamma_{mmmm})$,
or combinations of both strategies. Moreover, deviations of the same order of magnitude were obtained 
for the corresponding average transfer times (results not shown). \\
\indent Overall, this analysis seems to suggest that high transport efficiency in a tight-binding model 
with site disorder can be achieved robustly through noise, rather insensitively of the specific 
connectivity pattern of the tight-binding network. At the same time, one should be warned that efficiency 
is a rather {\em global} indicator. Therefore, some caution is required in interpreting this result as a thorough dismissal 
of tight and specific connections between the physical and geometrical structures underlying noise correlations and 
the propagation of quantum excitations coupled to the vibrating, noise-producing scaffold. 
Moreover, disregarding explicitly certain slow frequency components in the spectral densities, in accordance with
the white noise prescription, could prove a crucial limitation in the quest for quantitative 
links between protein vibrations and exciton transport. All in all, although descriptions of the 
HSR type prove that a certain degree of quantum coherence increases quantum efficiency over purely
incoherent, hopping transport, one should probably couple these with quantum-classical hybrid methods,
where structure-encoded spatiotemporal correlations provide an {\em explicit} description of noise. 
In the next section, based on the spectral features emerging from QM/MM correlations in the PE5454 complex, 
we develop a general method to dig for possible exciton-coupled vibrational patterns of the protein 
matrix.

%========================================================================================================================
\section{HSR coefficients: an elastic-network analysis reveals specific low-frequency normal modes}\label{sec:res_NM}

In this section, we consider the question whether the slow-frequency components emerging in certain 
QM/MM-based spectral densities (see again Fig.~\ref{f:sel_spectr}) flag specific {\em collective
vibrations}, whose pattern is encoded in the protein scaffold and that would selectively couple to 
exciton transport at the limits of validity of the HSR approach. \\
\indent The elastic-network model (ENM) denotes a class of  powerful and rather inexpensive 
tools that are well suited to  the investigation of collective fluctuations of 
proteins~\cite{Atilgan:2001zp,Hinsen:2000uq,Micheletti:2002fv,Tama:2002nv,Tirion:1996jw}.
In the residue coarse-grained (CG) version, also known as the anisotropic network model (ANM)~\cite{Atilgan:2001zp},
the protein is described as a network of as many fictive beads as there are amino acids. In the 
simplest case, each bead possesses the same mass equal to the average amino acid mass (about 110 a.m.u.) and
occupies at rest the position of the corresponding
$\alpha$ Carbon along the backbone chain of the protein as determined from an experimentally solved 
structure, i.e. X-ray or NMR.
The beads interact through identical Hooke springs 
with other beads in the structure located within a specified cutoff distance $R_c$. According to 
this prescription, the total potential energy of such elastic network model 
reads~\footnote{See Ref.~\cite{Bahar:2010aa} for a thorough discussion of alternative schemes.} 
\be\label{e:ENM}
U = \frac{1}{2} k_2 \sum_{i>j=1}^N c_{ij}(r_{ij} - R_{ij})^2 
\ee
where $N$ is the number of residues, $R_{ij}$ and $r_{ij}$ the equilibrium and instantaneous 
distances between residues $i$ and $j$, respectively, and $k_2$ is the common spring stiffness.
The matrix $c_{ij}$ specifies the connectivity of the network, namely 
$c_{ij}=\{ 1 \, \text{for} \ R_{ij} \leq R_c | 0 \ \text{otherwise} \}$. 
Consensus choices for the two parameters of such a model are $k_2 = 5$ kcal/\AA$^2$ and 
$R_c = 10$ \AA~\cite{Juanico:2007yw}. The normal modes can be easily computed as the eigenvectors 
of the Hessian matrix of the potential energy function~\eqref{e:ENM}~\cite{Sanejouand:2013aa}.
With this choice of the parameters, the coarse-grained 
NM spectrum of PE545 (PDB id. \texttt{1XG0}) 
spans the frequency interval $[3.9-93.5]$ cm$^{-1}$\\
%
%
%%%%%%%%%%%%%%%%%%%%%%%%%%%%%%%%%%%%%%%%%%%%%%%%%%%%%%%%%%%%%%%%%%%%%%%%%%%%%%%%%%%%%%%%%%%%%%%%%%%%%%%%%%%%%%%%%%%%%%%%%%%%%%%%%%
\begin{figure}%[t!,clip]
\begin{center}
\includegraphics[width=1\textwidth,clip]{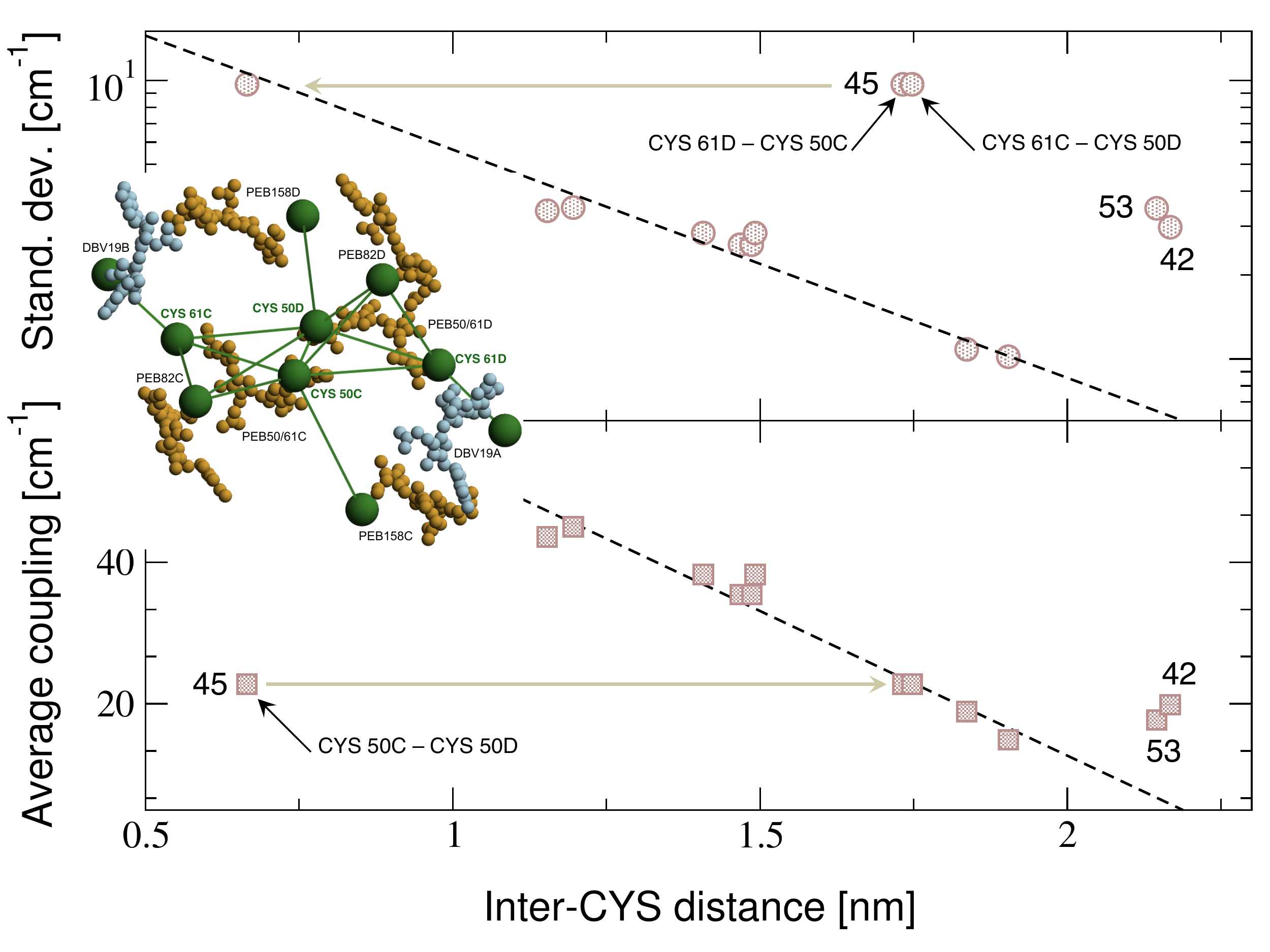}
\end{center}
\caption{Average absolute value (bottom) and standard deviation (top) of the QM/MM coupling time series versus
distance between the covalently attached cysteine residues. The dashed lines are plots of the apparent 
{\em typical} trends, namely the
exponential decay $\sigma(d) = \sigma_0 \exp(-d/d_\sigma)$, with $\sigma_0 = 37.21$ cm$^{-1}$, $d_\sigma = 5.3$ \AA \
(top) and $\langle J(d) \rangle = J_0 ( 1 - d/d_c)$, with  
$J_0 = 94.4$ cm$^{-1}$, $d_c = 23.1$ \AA. The cartoon shows the sub-network of covalently attached 
cysteine residues and the links flag the shortest 15 CYS-CYS inter-distances, from 
the closest pair, namely CYS 50C and CYS 50 D (6.65 \AA), to the pair CYS158C-CYS50C (21.67 \AA).
The couplings that appear to be outliers with anomalously large fluctuations and off-trend average couplings
are explicitly highlighted (the numbering is the same as the one used in Fig.~\ref{f:network}).
{\bf 53}: CYS 50D-CYS 158 D, {\bf 42}: CYS 50C-CYS 158D. 
}
\label{f:averCsigmdCYS}
\end{figure}
%%%%%%%%%%%%%%%%%%%%%%%%%%%%%%%%%%%%%%%%%%%%%%%%%%%%%%%%%%%%%%%%%%%%%%%%%%%%%%%%%%%%%%%%%%%%%%%%%%%%%%%%%%%%%%%%%%%%%%%%%%%%%%%%%%
%
\indent All chromophores in the PE5454 complex are covalently linked through disulfide bridges to 
specific cysteine residues (CYS)~\cite{doust2004}. The central pigments PEB50/61C and PEB50/61D form two disulfide
bridges each, while all the other bilins are covalently linked to one cysteine residue each. Overall,
these residues form an extended subnetwork of anchor points, whose concerted motions are likely to 
influence the coupling and site energy fluctuations of the attached pigments. It is instructive to
investigate how the average and the standard deviation of the QM/MM coupling time series depend upon 
the distance between the respective covalently linked CYS pairs. The data reported in Fig.~\ref{f:averCsigmdCYS}
suggest that on average both indicators referring to a given pair of pigments decrease with the distance between
the associated CYS anchor points (dashed lines). Interestingly, pairs associated with slow frequency components 
in the corresponding spectral densities (see Fig.~\ref{f:sel_spectr}) stand out from the common average 
trend as outliers. More precisely, fluctuations (gauged here by the standard deviations) of the 53 and 42 pairs 
are seen to be anomalously large as compared to the average trend at the same CYS-CYS distances. This is all the more 
remarkable as it concerns pairs located at large distances.
Another interesting observation is that the average coupling of the central PEB pair (45) appears 
to be on the common trend only when associated with pairs of opposite cysteines
(61C-50D and 61D-50C, see again the cartoon and the arrow in the bottom panel of Fig.~\ref{f:averCsigmdCYS}).
However, this is no longer true for the corresponding fluctuation, which appears to be of the correct 
order of magnitude only when associated with the two central cysteines (50C-50D, top horizontal arrow).
Overall, this suggests that the fluctuations in the 54 coupling are controlled by the relative motions
of the CYS50C-CYS50D pair. In turn, this points to the interaction between the associated terminal pyrrole rings 
(nearly parallel and about 5 \AA \ away from each other) as the main structural unit controlling the 
fluctuations of the 45 coupling strength.\\
\indent The above observations strongly suggests that there exist a subset of normal modes in the $30-80$ cm$^{-1}$
range whose displacement field (pattern) represents specific concerted motions of the cystein subnetwork.
Such modes, showing up in the spectral density referring to specific coupling-coupling and coupling-site energy 
(cross) correlations, are good candidates in the quest for fold-encoded, transport-enhancing vibrations.
In order to identify these modes, we have proceeded as follows. First, we have ranked normal modes in 
descending order with respect to their involvement fraction on the CYS subnetwork. Let us denote the 
latter as $\mathcal{S}_{\rm cys}$ and let $\vec{\xi}^{\,k}_i$ denote the displacement of particle (residue) $i$
in the mode $k$. NMs patterns are normalized, i.e. $\sum_{i=1}^{3N-6} |\vec{\xi}^{\,k}_i|^2=1$ $\forall \ k$,
so that the quantity 
\be
\label{e:fractionNM}
f_k =  \sum_{i\in \mathcal{S}_{\rm cys}} |\vec{\xi}^{\,k}_i|^2 \leq 1
\ee
gauges the relative weight of the covalently linked cysteines in the overall displacement field
of the $k$-th NM. If only those residues were vibrating in a given mode $m$, one would have $f_m = 1$.\\
\indent It would be tempting to rank all normal modes in descending order with respect to the 
associated $f_k$ values and select the top-scoring ones. However, a moment's thought is enough to 
realize that one also needs to gauge the {\em statistical significance} of a given $f_k$ score.
This should be done by considering for each mode the null hypothesis that a totally random subset 
of as many residues would give the same score. Thus, for each NM we have computed a 
$p$-value by performing a large number $N_R = 10^4$ of random reshufflings of the NM pattern and 
computing the corresponding value of the involvement score $f^R_k(i), i=1,2,\dots,N_R$. 
The statistical significance of the measured involvement for a given NM can then be estimated by 
the fraction of random scores exceeding the measured one. Namely,
\be
\label{e:pvalueINV}
p_f = \frac{1}{N_R}\sum_{i=1}^{N_R} \Theta\left( f^R_k(i) - f_k \right)
\ee
where $\Theta(x)$ is Heaviside function.
The quantity $1-p_f$ gauges the rejection probability of the null hypothesis. If we restrict to 
confidence levels greater than 95 \%, this analysis singles out 78 NMs with frequencies lying between 
$\omega_1 = 31.98$ cm$^{-1}$ and $\omega_2 = 77.49$ cm$^{-1}$, i.e. 6.9~\% of the NMs in 
the same frequency range and 5.3~\% of the total number of NMs. We term this ensemble $\mathcal{M}_{95}$.
Interestingly, this is precisely the same frequency
range where low-frequency structures appear in specific spectral-densities (see again Fig.~\ref{f:sel_spectr}).
It is possible to refine further the choice of possible interesting NMs by isolating vibrations whose 
vibrational patterns are mostly directed along the directions joining pairs of 
cysteines in the covalent subnetwork. We isolated the closest 15 pairs, which correspond 
to two links in the connectivity graph ($R_c=10$ \AA, inter-distances shorter than 21.67 \AA,
see cartoon in Fig.~\ref{f:averCsigmdCYS}). For each NM we computed the projection 
\be
%\begin{aligned}
\label{e:proj}
%&\phi_{ij}^k = \vec{\xi}^{\,k}_i \cdot \hat{R}_{ij} \\
%&\psi_{ij}^k = \vec{\xi}^{\,k}_j \cdot \hat{R}_{ij} \\
\mu_{ij}^k  = \frac{\vec{\xi}^{\,k}_i - \vec{\xi}^{\,k}_j}{\left|\left|\vec{\xi}^{\,k}_i - \vec{\xi}^{\,k}_j\right|\right|}
              \cdot \hat{R}_{ij}
%\end{aligned}
\ee
%
%
%++++++++++++++++++++++++++++++++++++++++++++++++++++++++++++++++++++++++++++++++++++++++++++++++++++++++++++++++++++++++++++
\begin{table}[t!]
\caption{\label{t:NM_notev} Values of $\mu_{ij}$ for different pairs in the CYS-CYS subnetwork
for the 8 normal modes with $p_f,p_\mu< 0.05$ The figure $\overline{\mu}$
is the average value over the considered 15 pairs (see text). The last
column reports the $\mu$-confidence level (c.l.), i.e. $1-p_\mu$. Normal modes 
are numbered starting from the highest frequency one.}
%\begin{tabular*}{\textwidth}{@{\extracolsep{\fill}} cc  ccccccc}
\begin{tabular}{cc | ccccccc}
\br
NM &  $\omega$ [cm$^{-1}$] 
   &  $\overline{\mu}$  &  50C-50D    &  61D-50C  
   &  50D-61C           &  158D-50D   &  158C-50C 
   &  c.l.  \\
\mr\mr
144  & 77.49  &  0.639 & -0.915 & -0.750 &  0.888 &  0.251 &  0.386  & 0.974  \\
245  & 72.45  &  0.691 &  0.866 &  1.000 & -0.659 &  0.215 & -0.974  & 0.995  \\
432  & 65.74  &  0.636 & -0.706 &  0.475 & -0.770 & -0.380 & -0.585  & 0.972  \\ 
501  & 63.37  &  0.639 &  0.900 & -0.901 & -0.478 &  0.904 &  0.750  & 0.971  \\
531  & 62.39  &  0.635 &  0.731 &  0.994 & -0.853 & -0.605 &  0.146  & 0.963  \\
546  & 61.87  &  0.764 & -0.991 &  0.317 &  0.634 & -0.980 &  0.850  & 1.000  \\  
556  & 61.50  &  0.650 &  0.402 & -0.814 &  0.965 & -0.505 &  0.652  & 0.976  \\
940  & 48.57  &  0.623 &  0.941 &  0.104 &  0.904 &  0.927 & -0.669  & 0.953  \\ 
\br
\end{tabular}
\end{table}
%++++++++++++++++++++++++++++++++++++++++++++++++++++++++++++++++++++++++++++++++++++++++++++++++++++++++++++++++++++++++
%
where $\hat{R}_{ij}$ denotes the unit vector joining cysteine $i$ to cysteine $j$ in the 
equilibrium structure. Again, one needs to evaluate the rejection probability of the null hypothesis 
stipulating that a value of $\mu$ equal or higher than a measured one 
can be obtained by selecting 15 random pairs of residues 
in the entire structure. We then considered the NMs in the previously identified
subset $\mathcal{M}_{95}$ and computed the indicators $\mu_{ij}^k$ and their associated $p$-values,
defined as 
\be
\label{e:pvalueproj}
p_\mu = \frac{1}{N_R}\sum_{i=1}^{N_R} \Theta\left( \overline{\mu}^R(i) - \overline{\mu}^k \right)
\ee
where $\overline{\mu}$ refers to the average of the projections~\eqref{e:proj} computed over 
the 15 closest cysteine pairs ($\overline{\mu}^k$) and over as many randomly selected pairs 
($\overline{\mu}^R$) for a given normal mode $k$ in the subset $\mathcal{M}_{95}$.\\
\indent Remarkably, 10~\% of these modes (8 NMs out of 78) showed a confidence level $1-p_\mu$ 
greater than 95~\%. A summary of this analysis is reported in Table~\ref{t:NM_notev}.
Such modes have frequencies in a somewhat narrower range than the previous 78 ones, lying
in the range [48.57,77.49] cm$^{-1}$ (see vertical lines in Fig.~\ref{f:sel_spectr}). The average 
projection on the first 15 CYS-CYS inter-distances over the whole subset is 0.66, suggesting 
that these NMs are indeed concerted motions of the CYS-CYS subnetwork, whose vibrational patterns 
can be imagined to a great extent as acting along the links of the associated connectivity 
subgraph (see green sticks in the cartoon in Fig.~\ref{f:averCsigmdCYS}). Remarkably, the average 
projection $\mu_{ij}$ on the 50C-50D direction (corresponding to the largest QM/MM coupling fluctuation) 
is 0.81, larger than the average projection computed for all the 15 pairs. This strongly 
confirms our previous inference about the crucial role of these two cysteines in mediating the 
protein-encoded fluctuations that couple to exciton transport.

%========================================================================================================================
\section{Conclusions and perspectives}

In this paper we have provided a realistic parametrization of the Haken-Strobl model~(\ref{eq:Silbey}) for the PE545
bilin complex from QM/MM simulations. The aim of our analysis was to dig for the different 
auto- and cross-correlations relating to pigment site energies and pigment-pigment couplings and 
investigate their role on exciton mobility.
We have used this approach to characterize exciton transport by means of simple global observables, namely
the transport efficiency and the average transfer time.
As a result, we have obtained that at room temperature the system is very close to the optimal transport regime
with a predominant contribution from site-energy fluctuations. Moreover, in the optimal regime the transport efficiency 
is rather insensitive to random reshuffling of the elements of the Hamiltonian and the associated noise tensor.
This suggests that noise-induced depinning of quantum excitations in disordered environments is a rather general 
feature of open quantum systems.   \\
\indent The high temperature regime revealed that the small residual coupling fluctuations in the dynamics
suppress the quantum Zeno regime at temperatures $T\gtrsim 5\,T_r$, where the dynamics is essentially that 
of a classical random walk. 
Although this regime is not meaningful from a biological standpoint, since  proteins unfold at temperatures 
$T\gtrsim 1.2\,T_r$, our study provides a different  reading frame for noise-assisted 
quantum transport~\cite{rebentrost2009environment} in photosynthetic complexes. 
Namely, the effect of environment  fluctuations at room temperature is to  provide access 
to {\it the same} efficiency level that would pertain to an inaccessible
regime of classical diffusion at extremely high temperatures. To put it in more evocative terms, 
one might argue that evolution has taken advantage of the laws of quantum mechanics to lower 
the effective temperature corresponding to the highest efficiency (the classical regime) to the region 
of working temperatures of proteins in a living organism.  \\
\indent We have shown that QM/MM data spotlight clear slow frequency components in the [30-80] cm$^{-1}$
range in the spectral densities associated with specific coupling-coupling and coupling-sit energy 
fluctuations. Typically these involve the central PEB50/61D/C pigments. 
These lie at the threshold of validity of the white noise approximation inherent 
in HSR-type approaches, where one has to surmise that the noise spectral structure does not
contain slow components, relaxing in time as fast as or more slowly than the timescale associated 
with exciton transfer.  \\
\indent While there seems to be no obvious way to include slow, collective modes in HSR-kind descriptions of 
exciton transport, in the last section of the paper we lay out a possible strategy to tackle this problem through
the elastic-network model formalism.  Our analysis clearly identify a set of specific collective normal modes 
that preferentially couple the subnetwork of 10 cysteine residues covalently linked to bilins.
Such modes, selected through an original double statistical-significance analysis, have the combined properties of 
(i) maximum concerted involvement on the CYS subnetwork and (ii) maximum displacement {\em alignment} with the 
ensemble of the closest inter-distance vectors between cysteine pairs. These modes, with frequency 
lying in the [50-80] cm$^{-1}$ range, might be important fold-encoded {\em noise} sources for 
enhancing exciton transport optimally within the pigment-hosting protein matrix. \\
\indent The logical step beyond microscopically parameterized HSR approaches to investigate 
the above hypothesis is to turn to coarse-grained  quantum-classical hybrid descriptions, 
often also called mixed quantum classical approaches (QCA)~\cite{May:2011aa}.
In QCAs the force acting on classical degrees of freedom is computed by considering an effective 
potential energy given by the quantum mechanical expectation value of
the system-reservoir coupling according to the instantaneous values of the classical
coordinates and the wave function of the quantum degrees of freedom (DOFs).
While the presence of the classical DOFs results in an additional potential for the dynamics of the quantum
system, the classical reservoir feels the quantum mechanically averaged force of the
relevant quantum system. This method is often referred to as Ehrenfest method~\cite{May:2011aa}
and falls within the class of mean-field approaches. 
Interestingly, a recent QCA based on the nonlinear network model~\cite{Juanico:2007yw} 
has shown that specific fold-rooted vibrational modes have the potential to spotlight 
alternative excitation energy transfer routes in the 
Fenna-Matthews-Olson (FMO) complex through their influence on pigment properties~\cite{Morgan:2016aa}.
Overall, also in view of the results of the previous section, elastic/nonlinear-network model-based 
QCAs can be regarded as optimal tools to investigate the role of
specific fold-encoded vibrational modes in exciton transport 
in light-harvesting complexes.

%%%%%%%%%%%%%%%%%%%%%%%%%%%%%%%%%%%%%%%%%%%%%%%%%%%%%%%%%%%%%%%%%%%%%%%%%%%%%%%%%%%%%%%%%%%%%%%%%%%%%%%%%%%%%%%%%%%%%%%%%%%%%%%

\bigskip
\section*{Acknowledgments}
S.P., S.I., Y.O. and  F.P. thank financial support from the EU FP7 project PAPETS (GA 323901).
S.P. and Y.O. thank the support from Funda\c{c}\~{a}o para a Ci\^{e}ncia e a Tecnologia (Portugal), 
namely through programmes PTDC/POPH/POCH and projects UID/EEA/50008/2013, IT/QuSim, IT/QuNet, ProQuNet, 
partially funded by EU FEDER, as well as project CRUP/CPU CQVibes TC16/14.
Furthermore S.P. acknowledges the support from the DP-PMI and FCT (Portugal) through scholarship PD/BD/52549/2014.

\bigskip
\section*{References}
%%%%%%%%%%%%%%%%%%%%%%%%%%%%%%%%%%%%%%%%%%%%%%%%%%%%%%%%%%%%%%%%%%%%%%%%%%%%%%%%%%%%%%%
%\bibliography{biblio_bilins,bio-11} 
%%%%%%%%%%%%%%%%%%%%%%%%%%%%%%%%%%%%%%%%%%%%%%%%%%%%%%%%%%%%%%%%%%%%%%%%%%%%%%%%%%%%%%%

\providecommand{\newblock}{}

\end{document}

\title{Supporting information}

\maketitle

%%%%%%%%%% Merge with supplemental materials %%%%%%%%%%
%%%%%%%%%% Prefix a "S" to all equations, figures, tables and reset the counter %%%%%%%%%%

\setcounter{equation}{0}
\setcounter{figure}{0}
\setcounter{table}{0}
\setcounter{page}{1}
\makeatletter
\renewcommand{\theequation}{S\arabic{equation}}
\renewcommand{\thefigure}{S\arabic{figure}}
\renewcommand{\bibnumfmt}[1]{[S#1]}
\renewcommand{\citenumfont}[1]{S#1}

\section{Exciton Hamiltonian parameters}

\begin{table}[h]
 \begin{tabular}{|l|c|c|c|c|c|c|c|c|}
 \hline
 &     $DBV_{19A}$& $DBV_{19B}$& $PEB_{158C}$& $PEB_{158D}$& $PEB_{50/61C}$& $PEB_{50/61D}$ &$PEB_{82C}$& $PEB_{82D}$\\
 \hline
$DBV_{19A}$ & -  &   -4.3 &   -27.3  &    3.5   &   2.2  &  -39.3  &  -11.4  &   34.3 \\
   \hline
$DBV_{19B}$ &-4.3 &     -  &  -3.7  &   26.3  &  -42.6   &   1.4  &  -36.1   &  11.6\\
  \hline  
$PEB_{158C}$& -27.3  &   -3.7   &   -  &  -6.1  &  -21.5  &  -15.2  &    7.3    &  6.4\\
 \hline
$PEB_{158D}$ & 3.5   &  26.3   &  -6.1    &  -  &   24.5 &    19.1   &   6.8    &  8.2 \\ 
   \hline   
$PEB_{50/61C}$&  2.2   & -42.6 &   -21.5  &   24.5   & -    &  71.7    & 34.0 &    12.1\\
   \hline 
$PEB_{50/61D}$ &-39.3  &    1.4 &   -15.2    & 19.1  &   71.7   &  -  &  -16.0  &  -35.6\\
 \hline
$PEB_{82C}$ &-11.4  &  -36.1    &  7.3    &  6.8   &  34.0    &-16.0  &    - &     4.0\\
\hline
$PEB_{82D}$ & 34.3   &  11.6     & 6.4  &    8.2  &   12.1   & -35.6   &   4.0    &  - \\
  \hline
 \end{tabular}
 \caption{MD averaged electronic couplings in cm$^{-1}$, see Ref.~\cite{curutchet2013}}
 \label{tab:couplings}
\end{table}

\begin{table}[h]
 \begin{tabular}{|l|c|c|c|c|c|c|c|c|}
 \hline
 &     $DBV_{19A}$& $DBV_{19B}$& $PEB_{158C}$& $PEB_{158D}$& $PEB_{50/61C}$& $PEB_{50/61D}$ &$PEB_{82C}$& $PEB_{82D}$\\
 \hline
 SET 1& 277& 0.0& 1014 &817 &1531& 1305& 544& 544\\
 \hline
 SET 2& 0.& 0.& 800 &650 &1450& 1050& 550& 50\\
 \hline
 \end{tabular}
 \caption{Pigment site energies (in cm$^{-1}$) reported in Ref.~\cite{curutchet2013}. 
 Site-energies in Set 1 are directly extracted from QM/MM simulations. Parameters in Set 2 are further adjusted
 in order to better fit experimental measurements.}
  \label{tab:SE}
\end{table} 
The Hamiltonian $H_0$ referred to in the main text was obtained by combining 
the couplings in Tab.~\ref{tab:couplings} with parameter
set 1 in Tab.~\ref{tab:SE}. In Fig.~\ref{f:efficiency_supp} we show the same study as in Fig. 4 
repeated for the Hamiltonian $\tilde H_0$, which  was obtained by combining the 
couplings in Tab.~\ref{tab:couplings} with parameter set 2 in Tab.~\ref{tab:SE}.

%%%%%%%%%%%%%%%%%%%%%%%%%%%%%%%%%%%%%%%%%%%%%%%%%%%%%%%%%%%%%%%%%%%%%%%%%%%%%%%%%%%%%%%%%%%%%%%%%%%%%%%%%%%%%%
\section{Environmental parameters}

The tensor $\Gamma_{mnpq}$ was computed from QM/MM trajectories of the time-dependent exciton Hamiltonian
$H(t)$ (see Eq.~(2)) for the bilin complex. 
%
%{\bf details of the QM/MM setup}.
%
After a suitably long transient, each trajectory was sampled with a time step $\delta t=5$ fs for a total time
$t_f = 300$ ps. The correlation function 
\be
C_{mnpq}(\tau)=\left\langle H_{mn}(t)H_{pq}(t') \right\rangle -\left\langle 
H_{mn}(t)\right\rangle \left\langle H_{pq}(t') \right\rangle
\ee
was then computed by averaging over moving temporal windows with total length $5$ ps spaced by $0.1$ ps.
A Lorentzian fit of the cosine transform $C_{mnpq}(\omega)$ of the form $f(\omega)=a/(\omega^2+\omega_c^2)$ 
was then performed in the low frequency spectral region $\omega \in [0, 500]$ cm$^{-1}$.
The coupling coefficients $\Gamma_{mnpq}$ were computed as the zero-frequency limit of $f(\omega)$, i.e. $\Gamma=f(0)=a/\omega_c^2$.
Finally, we have neglected the terms $\Gamma_{mnpq}$ corresponding to spectral widths $\omega_c< 2\pi/\Theta$, where the threshold
$\Theta$ was set to $1$ ps, see the main text.
The result for the environment tensor  $\Gamma_{mnpq}$ is shown in Fig. 3. 
Here we explicitly report our parametrization of  $\Gamma_{mnpq}$ in the limit of vanishing
cross-correlations, i.e.
\be
\Gamma_{mnpq}=\left[\delta_{mp}\delta_{nq}+\delta_{mq}\delta_{np}(1-\delta_{mn})   \right]\gamma_{mn}
\label{eq:approx}
\ee
The reduced matrix $\gamma_{mn}$ is shown in Tab.~\ref{tab:redc_gamma} 
\bgroup
\def\arraystretch{1.5}
\begin{table}[h]
 \begin{tabular}{|c|c|c|c|c|c|c|c|c|}
 \hline
index &0& 1& 2& 3& 4& 5& 6& 7\\ 
\hline
0& 105.5&   0&      0.046 &       0 &      0.007 &      0.028 &       0.002 &       0.032 \\
 \hline
1& 0&    104.8&     0 &       0.061 &      0.026 &      0 &	0.034 &	 0.002 \\
 \hline
2& 0.046&    0&       127.8 &     0 &      0.017 &      0.014 &	0.001 &	 0.002\\
 \hline
3& 0&    0.061&       0 &       123 &    0.026 &      0.020 &	0.019 &	 0.001 \\
 \hline
4& 0.007&    0.026&       0.017 &       0.025 &      83.4 &    0.211 &	0.010 &	 0.001 \\
 \hline
5& 0.028&    0&       0.014 &       0.020 &      0.211 &      89.6 &	0.002 &	 0.015 \\
 \hline
6& 0.002&    0.034&       0.001&       0.002 &      0.010 &      0.002 &	66.8 &	 0 \\
 \hline
7& 0.033&    0.002&       0.002 &       0.001 &      0.001 &      0.015 &	0 &	 60.1\\
 \hline
 \end{tabular}
 \caption{Reduced coupling matrix $\gamma_{mn}$ in cm$^{-1}$  computed from the QM/MM trajectories.}
 \label{tab:redc_gamma}
\end{table} 
\egroup
\clearpage

%%%%%%%%%%%%%%%%%%%%%%%%%%%%%%%%%%%%%%%%%%%%%%%%%%%%%%%%%%%%%%%%%%%%%%%%%%%%%%%%%%%%%%%%%%%%%%%%%%%%%%%%%%%%%%%%%%
\section{site energies versus couplings  correlations}

\begin{figure}[ht!,clip]
\begin{center}
\includegraphics[width=0.49\textwidth]{SM_pigment0.eps}
\includegraphics[width=0.49\textwidth]{SM_pigment1.eps}
\includegraphics[width=0.49\textwidth]{SM_pigment2.eps}
\includegraphics[width=0.49\textwidth]{SM_pigment3.eps}
\includegraphics[width=0.49\textwidth]{SM_pigment4.eps}
\includegraphics[width=0.49\textwidth]{SM_pigment5.eps}
\includegraphics[width=0.49\textwidth]{SM_pigment6.eps}
\includegraphics[width=0.49\textwidth]{SM_pigment7.eps}
\end{center}
\caption{Coupling coefficients $\Gamma_{mmpq}$  in sector II (site energy - couplings correlations). 
Each panel shows the amplitdues involving a fixed site energy $m$  and all the possible couplings $pq$. Please
notice that the vertical scale is pigment-dependent. Data refer to a threshold $\Theta=1$  ps.} 
\label{f:Gamma_tensor}
\end{figure}
 
\section{Exciton transfer efficiency} 

\begin{figure}[ht,clip]
\begin{center}
\includegraphics[width=0.8\textwidth]{efficiencyH1.eps}
\end{center}
\caption{
The same as Fig.4, but with exciton Hamiltonian $\tilde H_0$ instead of $H_0$.
}
\label{f:efficiency_supp}
\end{figure}

\bibliography{biblio}
\bibliographystyle{apsrev}